\documentclass[useAMS,usenatbib]{mn2e}

\usepackage{graphicx}
\usepackage{grffile}\usepackage{epsf}
\usepackage{amsmath}
\usepackage{float}
\usepackage{fixltx2e}

\newcommand{\kmsmpc}{\kms\;{\rm Mpc}^{-1}}

\newcommand{\kms}{{\rm km}\,{\rm s}^{-1}}
\newcommand{\cms}{{\rm cm}^{-2}}
\newcommand{\cmc}{{\rm cm}^{-3}}

\newcommand{\msolar}{{\rm M}_{\odot}}
\newcommand{\msolaryr}{{\rm M}_{\odot} {\rm yr}^{-1}}

\newcommand{\gad}{{\sc Gadget-3}}

\newcommand{\OI}{\hbox{O\,{\sc i}}}

\newcommand{\OIV}{\hbox{O\,{\sc iv}}}

\newcommand{\OVI}{\hbox{O\,{\sc vi}}}
\newcommand{\OVII}{\hbox{O\,{\sc vii}}}
\newcommand{\OVIII}{\hbox{O\,{\sc viii}}}
\newcommand{\Oxy}{\hbox{O}}
\newcommand{\OIX}{\hbox{O\,{\sc ix}}}
\newcommand{\HI}{{\hbox{H\,{\sc i}}}}

\newcommand{\nh}{{n_{\rm H}}}

\newcommand{\apj}{{ApJ}}

\begin{document}
\title[Bimodality of circumgalactic O VI]{Bimodality of low-redshift circumgalactic O VI in non-equilibrium EAGLE zoom simulations}

\author[B. D. Oppenheimer et al.]{
\parbox[t]{\textwidth}{\vspace{-1cm}
Benjamin D. Oppenheimer$^{1}$\thanks{benjamin.oppenheimer@colorado.edu}, Robert A. Crain$^{2}$,  Joop~Schaye$^{3}$, Alireza~Rahmati$^{4}$, Alexander J. Richings$^{5}$, James W. Trayford$^{6}$, Jason Tumlinson$^{7}$, Richard G. Bower$^{6}$, Matthieu Schaller$^{6}$, Tom Theuns$^{6}$}\\\\  
$^1$CASA, Department of Astrophysical and Planetary Sciences, University of Colorado, 389 UCB, Boulder, CO 80309, USA\\
$^2$Astrophysics Research Institute, Liverpool John Moores University, 146 Brownlow Hill, Liverpool, L3 5RF, UK\\
$^3$Leiden Observatory, Leiden University, P.O. Box 9513, 2300 RA, Leiden, The Netherlands\\
$^4$Institute for Computational Science, University of Z\"urich, Winterthurerstrasse 190, CH-8057 Z\"urich, Switzerland\\	
$^5$Department of Physics and Astronomy and CIERA, Northwestern University, 2145 Sheridan Road, Evanston, IL 60208, USA\\
$^6$Institute for Computational Cosmology, Durham University, South Road, Durham, DH1 3LE, UK\\
$^7$Space Telescope Science Institute, 3700 San Martin Drive, Baltimore, MD, USA\\
}
\maketitle

\pubyear{2016}

\maketitle

\label{firstpage}

\begin{abstract}

We introduce a series of 20 cosmological hydrodynamical simulations of
$L^*$ ($M_{200}=10^{11.7}-10^{12.3} \msolar$) and group-sized
($M_{200}=10^{12.7}-10^{13.3} \msolar$) haloes run with the model used
for the EAGLE project, which additionally includes a non-equilibrium
ionization and cooling module that follows 136 ions.  The simulations
reproduce the observed correlation, revealed by COS-Halos at $z \sim
0.2$, between $\OVI$ column density at impact parameters $b < 150$ kpc
and the specific star formation rate (sSFR$\equiv$SFR$/M_*$) of the
central galaxy at $z\sim 0.2$.  We find that the column density of
circumgalactic $\OVI$ is maximal in the haloes associated with $L^*$
galaxies, because their virial temperatures are close to the
temperature at which the ionization fraction of $\OVI$ peaks ($T\sim
10^{5.5}$ K).  The higher virial temperature of group haloes ($>10^6$
K) promotes oxygen to higher ionization states, suppressing the $\OVI$
column density.  The observed $N_{\OVI}$-sSFR correlation therefore
does not imply a causal link, but reflects the changing characteristic
ionization state of oxygen as halo mass is increased.  In spite of the
mass-dependence of the oxygen ionization state, the most abundant
circumgalactic oxygen ion in both $L^*$ and group haloes is $\OVII$;
$\OVI$ accounts for only 0.1\% of the oxygen in group haloes and
0.9-1.3\% with $L^*$ haloes.  Nonetheless, the metals traced by $\OVI$
absorbers represent a fossil record of the feedback history of
galaxies over a Hubble time; their characteristic epoch of ejection
corresponds to $z > 1$ and much of the ejected metal mass resides
beyond the virial radius of galaxies. For both $L^*$ and group
galaxies, more of the oxygen produced and released by stars resides in
the circumgalactic medium (within twice the virial radius) than in the
stars and ISM of the galaxy.

\end{abstract}


\section{Introduction}  

A self-consistent model of galaxy formation requires an understanding
of the relationship between the galactic components, stars and the
interstellar medium (ISM), and the surrounding gaseous reservoir that
extends to the virial radius and beyond, defined roughly as the
circumgalactic medium (CGM).  At redshift $z<0.5$, the galactic
component is well characterised through observational inference of
stellar masses, star formation rates (SFRs), and morphologies.  The
corresponding CGM remains much more challenging to characterise,
because its diffuse nature generally requires absorption line
spectroscopy toward background UV-bright sources.  This significant
reservoir of halo gas clearly plays a central role in the cycling of
baryons into and out of a galaxy, regulating its growth; however the
details of the galaxy-CGM relationship remain elusive.

Until recently, UV absorption line spectroscopy could not target
low-redshift galactic haloes, because only a small number of quasars
were bright enough.  The study of the metal enrichment of the low-$z$
intergalactic medium (IGM) relied on surveys of metal lines in
serendipitous sight lines providing a cross section-weighted sampling
\citep[e.g.][]{tho08b,tri08,wak09,til12,dan16}.  These observations
could be confronted with cosmological hydrodynamical simulations that
included metal enrichment from galactic superwinds to learn how the
nucleosynthetic products of star formation enrich the diffuse CGM and
IGM \citep[e.g.][]{agu01a, sch03, spr03b, opp06, opp09a, tor10, wie10,
  cen11, smi11, tep11, tes11, opp12, cra13, rah15b}.

The installation of the Cosmic Origins Spectrograph (COS) greatly
increased the throughput of UV spectroscopy on the {\it Hubble Space
  Telescope} (HST) platform \citep[e.g.][]{sav10}, enabling the use of
fainter quasar targets illuminating foreground galactic haloes.
Several HST programs have exploited this new capability to investigate
the gaseous environments of a sample of low-$z$ haloes, including
COS-Halos \citep{tum13}, COS-Dwarfs \citep{bor14}, and COS-GASS
\citep{bort15}.  Other recent studies have also characterised the CGM
relative to samples of galaxies, including \citet{che10, bor11, kac11,
  bort13, chu13, nie13, sto13, lia14, kac15} at low-$z$ and
\citet{tur14} at high-$z$.  

One of the first COS-Halos results demonstrated statistically that the
properties of $\OVI$ absorption, out to impact parameter $b=150$ kpc,
correlate with the properties of the host galaxies \citep[][hereafter
T11]{tum11}.  The strongest correlation was found between the $\OVI$
column density ($N_{\OVI}$) and the specific star formation rate
(sSFR$\equiv$SFR$/M_*$) of the galaxy, such that the bimodality in
sSFR between the COS-Halos star-forming sample and the passive sample
is clearly reflected in their $\OVI$ columns.  Of the 12 COS-Halos
passive galaxies, defined as having sSFR less than $10^{-11}$
yr$^{-1}$, at least half have $N_{\OVI}<10^{14.0} \cms$, while 90\% of
the 30 COS-Halos star-forming galaxies have $N_{\OVI}>10^{14.2} \cms$
and the median $N_{\OVI}$ is $10^{14.5} \cms$ at $b<150$ kpc.

T11 identified $\OVI$ haloes as a major reservoir of the metals
synthesized by star-forming galaxies, estimating a typical $\OVI$ mass
of $\approx 2\times 10^6 \msolar$ when summing the median $\OVI$
column of $10^{14.5} \cms$ over a 150 kpc radius aperture.  The total
mass of oxygen traced by COS-Halos is at least $1.2\times 10^7
\msolar$ when assuming a maximum $\OVI$ ionization fraction,
$x_{\OVI}=0.2$.  \citet{pee14} explored the $\OVI$ haloes in the
context of the total metal budget estimated to be synthesized by
galaxies, concluding that there exists nearly as much metal mass
traced by circumgalactic $\OVI$ as in the ISM of galaxies with
$M_{*}<10^{10} \msolar$ for $x_{\OVI}=0.2$.

The discovery of such a large reservoir of $\OVI$-traced metals
extending far beyond the stellar disk of the galaxy leads to several
questions about the dynamical link between galaxies and their
associated CGM.  Is there a direct relation between the $\OVI$ in
haloes and star formation as suggested by the $\OVI$-sSFR correlation?
Such a link would imply that typical $L^*$ galaxies are spewing
enriched gas in galactic winds out to 150 kpc and beyond in the
low-$z$ Universe.  A number of large-volume simulations
\citep[e.g.][]{for13, for15, sur15, sur16} and zooms with cosmological
initial conditions \citep[e.g.][]{sti12, hum13, she14, gut16} have
found an intimate link between galaxies and their metal-enriched CGM.
Simulations that specifically compared with the COS-Halos survey find
lower $\OVI$ column densities than observed around star-forming
galaxies \citep{for15,sur16}.  The \citet{for15} simulations do not
include Active Galactic Nucleus (AGN) feedback and do not reproduce
the $\OVI$ bimodality, while \citet{sur16} argue that AGN feedback
clears out the CGM around passive galaxies thus reducing $\OVI$ column
densities around passive galaxies.

Another possibility is that CGM $\OVI$ is not directly attributable to
recent star formation or AGN activity, and that instead the $\OVI$
bimodality arises from a more fundamental relationship.  To assess
this possibility, we need to ask what is the physical nature of
$\OVI$?  If $\OVI$ is predominantly photo-ionized gas at $\sim 10^4$ K
\citep[e.g.][]{opp09a}, then such absorbers could be a signature of
cold flows \citep[e.g.][]{ker05, dek09, van12} feeding star formation,
which may subside around passive galaxies.  Enhanced photo-ionization
from young stars could also ionize $\OVI$ around star-forming galaxies
\citep{opp13b, vas15, sur16}.  \citet{opp12} demonstrated that
increasing the ionization background strength at the $\OVI$ ionization
potential can result in photo-ionized $\OVI$ tracing higher densities,
even in circumgalactic halo gas.  Alternatively, if $\OVI$ is
primarily collisionally ionized at its $10^{5.5}$ K peak
\citep[e.g.][]{hec02, cen11, smi11, tep11, fae16}, then it signals a
hot halo component apparently associated with star-forming galaxies.
Is this warm-hot component related to recent outflows or does it trace
the temperature of a virialized, quasi-static halo?  If the latter, do
passive galaxies lack $\OVI$ haloes, because their virial temperatures
are much hotter than $10^{5.5}$ K?

Knowledge of the physical cause of the ionization is required to
constrain $x_{\OVI}$, from which the total mass of oxygen and metals
in the CGM is inferred.  Because the $\OVI$-traced CGM mass scales
inversely with $x_{\OVI}$, far more CGM oxygen would be present if
$x_{\OVI}$ is much lower than the maximal value implied by ionization
analyses.  Hence, it is uncertain how much underlying CGM oxygen the
COS-Halos $\OVI$ observation indicate.

To understand the origin and nature of $\OVI$ in a set of haloes
hosting galaxies corresponding to those probed by COS-Halos, we need
cosmological simulations that can simultaneously model the stellar
build-up of galaxies, the feedback processes that eject gas from
galaxies, and the detailed atomic processes that govern the abundances
of the ions observed in the CGM.  Here, we run a series of zoom
simulations targeting a range of galactic haloes similar to those
probed by the COS-Halos survey using the same model as the EAGLE
simulations \citep[][hereafter S15; Crain et al. 2015]{sch15}.  The
parameters of the subgrid physics routines employed by the EAGLE
simulations were calibrated to reproduce the observed $z\sim 0.1$
galactic stellar mass function, the super-massive black hole (SMBH)
mass-$M_*$ relation, and the sizes of galaxy discs using a combination
of star formation (SF)-driven thermal winds and SMBH thermal winds.
Observations of the CGM were not used to calibrate the EAGLE
simulations, therefore the outcomes here are genuine predictions of
the model.

Here we integrate the non-equilibrium (NEQ) ionization and dynamical
cooling module introduced by \citet[][hereafter OS13]{opp13a} into the
EAGLE simulation code to trace the evolution of 136 ions of 11
elements on the fly, which we activate at late times for evolved
haloes.  The module includes photo-ionization from a slowly evolving
meta-galactic background, in this case the \citet{haa01} quasar+galaxy
background.  We also note that this work is the first we know of to
integrate ion-by-ion cooling rates including for the metals
\citep{gna12,opp13a} into a cosmological simulation.  NEQ ionization
of oxygen was explored in simulations by \citet{cen06b} and
\citet{yos06}, but they did not perform dynamical cooling.  We discuss
the NEQ effects in Appendix \ref{sec:NEQ}, but note that these do not
significantly affect the $\OVI$ columns when using a uniform
meta-galactic background, nor do they alter the integrated properties
of galaxies ($M_{*}$, SFR, colour).

Our investigation focuses on the $\OVI$ column density observations
from COS-Halos.  We stress that the EAGLE feedback model has not been
calibrated to reproduce the properties of the CGM.  Nevertheless,
EAGLE shows agreement with cosmic $\HI$ statistics tracing the IGM
over a Hubble time \citep{rah15a}.  Furthermore, because EAGLE
reproduces key stellar and cold ISM properties of galaxies
\citep[e.g. S15;][]{fur15,fur16,lag15,lag16,tra15,bah16,seg16}, while
explicitly following the hydrodynamics, it is an ideal testbed for
this study and will indicate whether additional physical processes are
necessary to reproduce CGM measurements.  This enables us to consider
the total metal budget inside and beyond galaxies as done by
\citet{bou07, zah12, pee14}.  We also compare our results to the
related work of \citet{rah15b}, who explored IGM metal-line
statistics, including $\OVI$, in the EAGLE cosmological boxes.
Throughout we use the EAGLE-Recal prescription for all zooms as
discussed in S\ref{sec:sims}; however it should be noted that the main
EAGLE cosmological boxes use two different prescriptions (Ref and
Recal) at two different resolutions, which we we discuss in Appendix
\ref{sec:resconv} and show that there is not a significant difference
for our main results.

The layout of the paper is as follows.  We describe our zoom
simulations and implementation of our non-equilibrium module in
\S\ref{sec:sims}.  We present the physical nature and origin of the
oxygen-traced CGM in \S\ref{sec:phys}, and then confront the COS-Halos
observations in \S\ref{sec:COS-Halos}.  The global content of oxygen
produced and distributed by galaxies is considered in
\S\ref{sec:oxycont}.  We discuss the implications of our results in
\S\ref{sec:discuss} and summarize in \S\ref{sec:summary}.
Non-equilibrium effects and resolution convergence are explored in the
Appendix.

\section{Simulations} \label{sec:sims}  

In this Section, we give a brief overview of the EAGLE simulation
code, followed by how our non-equilibrium module is integrated with
it.  We then describe our zoom simulations with attention paid to how
the adopted physical model used here differs from the standard EAGLE
implementation.

\subsection{The EAGLE simulation code}

The EAGLE simulation code is an extensively modified version of the
N-body+Smoothed Particle Hydrodynamic (SPH) code \gad~last described
by \citet{spr05}, which is described in detail by S15.  The main
modifications to the entropy-conserving SPH implementation of the
\gad~code \citep{spr03a} comprise i) the pressure-entropy SPH
formulation described by \citet{hopk13}, ii) the artificial viscosity
switch of \citet{cul10}, iii) the artificial conduction switch of
\citet{pri08}, iv) the timestep limiter from \citet{dur12} to improve
energy conservation during sudden changes of internal energy, and v) a
C2 \citet{wen95} 58-neighbour kernel to suppress particle-pairing
instabilities \citep{deh12}.  Collectively, these updates are referred
to as ``Anarchy'' (Dalla Vecchia, in prep.; see also Appendix A of
S15) and the impact of their inclusion on the galaxy population was
explored by \citet{schal15}.

A number of subgrid physics modules are included in the EAGLE code.
Equilibrium radiative cooling and photo-heating rates are computed for
11 elements exposed to the \citet{haa01} extra-galactic ionization
background plus the Cosmic Background Radiation as described by
\citet{wie09a}.  These rates are computed using {\tt CLOUDY}
\citep[version 07.02][]{fer98}, and are replaced here with the NEQ
rates of OS13 described in \S\ref{sec:NEQcode} when run in NEQ mode.

Star formation is modelled using the \citet{sch08} pressure-based law.
A pressure floor, $P_{\rm EOS}\propto \rho^{4/3}$ is additionally
imposed, corresponding to a polytropic equation of state, normalised
to $T_{\rm EOS} = 8000$ K at $\nh = 0.1 \cmc$, because we do not
resolve the Jeans scales of the cold ISM.  The \citet{ken98} SF
surface density relationship is reproduced by construction for gas
particles above the \citet{sch04} metallicity-dependent SF density
threshold.  The stellar evolution and enrichment module introduced by
\citet{wie09b} includes element-by-element and mass metal loss from
AGB stars, stellar winds from massive stars, and SNe (core collapse
and Type Ia), which is distributed across the SPH kernel of each
stellar particle.  A \citet{cha03} initial mass function (IMF) is
adopted.

Following \citet{dal12}, feedback associated with star formation and
the SMBH growth is implemented by stochastically heating gas particles
neighbouring newly-formed star particles and accreting SMBH.  The
energy budgets available associated with stellar feedback and black
hole growth govern the probability of heating each SPH particle.
Briefly, stellar feedback heats particles by $\Delta T = 10^{7.5}$K in
a single episode 30 Myr after a star particle forms.  The fraction of
the available energy injected into the ISM depends on the local gas
metallicity ($Z$) to account for more efficient cooling at higher $Z$,
and the local gas density to compensate for artificial radiative
losses that occur at high density \citep[see S15;][for further
  motivation]{cra15}.  SMBH growth and AGN feedback are implemented
using a single thermal mode following \citet{boo09} with the addition
of the suppression of high angular momentum gas accretion following
\citet{ros13}.  We use the high-resolution EAGLE-Recal prescription,
which applies a heating temperature of $\Delta T_{\rm AGN} =
10^{9.0}$K.

\subsection{Non-equilibrium network integrated into EAGLE} \label{sec:NEQcode}
 
The NEQ module, introduced by OS13, explicitly follows the reaction
network of 136 ionization states of all 11 elements that contribute
significantly to the cooling (H, He, C, N, O, Ne, Si, Mg, S, Ca, \&
Fe) plus the electron density of the plasma.  These are the same
elements as are present in the equilibrium cooling module of
\citet{wie09a}, enabling a self-consistent switching between using
equilibrium lookup tables and the NEQ method.  Our reaction network,
described fully by OS13, includes radiative and di-electric
recombination, collisional ionization, photo-ionization, Auger
ionization, and charge transfer.  Cooling is performed ion-by-ion
\citep[][OS13]{gna12} summing over all 136 ions.  OS13 verified that
this method reproduces published results obtained with other codes.
We use particle-based (instead of kernel-smoothed) ion and metal
abundances for NEQ ionization and cooling as explained in
\S\ref{sec:runequil}.

Our NEQ module applies the Sundials {\tt
  CVODE}\footnote{https://computation.llnl.gov/casc/sundials/main.html}
solver to integrate the ionization and cooling over a hydrodynamic
timestep using the backward difference formula and Newton iteration.
Although an extended version of the network is also capable of
modelling ISM chemistry including molecular formation, cooling, and
dissociation \citep[see][]{ric14a}, we focus on the CGM and do not use
the NEQ network for gas densities above the metallicity-dependent SF
threshold.  For ISM gas, defined as having non-zero SFR, we use
equilibrium lookup tables tabulated as functions of density assuming
$T=10^4$ K, although we note that this matters little as the ion
abundances will rapidly evolve when the SPH particle is heated by
feedback, and the NEQ network followed.  Upon enrichment, SPH
particles receive the new metals in their ground-state ions.  The vast
majority of enrichment occurs in the ISM gas, where the network is not
used.  However, enrichment of gas followed using the NEQ network
occurs when stars outside of galaxies enrich via delayed feedback
(e.g. Type Ia SNe \& AGB winds).  Our exploration here focuses on CGM
gas traced by oxygen absorption, most of which is far from galaxies;
hence the details of the ion abundances in ISM gas and from recent
enrichment is essentially inconsequential.  We set the {\tt CVODE}
absolute tolerance to $10^{-9}$ and the relative tolerances to
$10^{-3}$.  We do not use a self-shielding prescription for dense gas
in either equilibrium or NEQ.  

\subsection{Zoom simulations}

\subsubsection{Initial conditions}

We assume the same \citet{pla14} cosmological parameters adopted by
the EAGLE simulations: $\Omega_{\rm m}=0.307$,
$\Omega_{\Lambda}=0.693$, $\Omega_b=0.04825$, $H_0= 67.77$ $\kmsmpc$,
$\sigma_8=0.8288$, and $n_{\rm s}=0.9611$.  To generate the zoomed
initial conditions, we use initial conditions generated using the
second-order Lagrangian perturbation theory method of \cite{jen10} and
the public Gaussian white noise field \textsc{Panphasia}
\citep{jen13a}.  Targets for resimulation are identified at $z=0$, and
the particles comprising a spherical region that encloses the target
object to a radius of $3 R_{200}$ (where $R_{200}$ encloses an
overdensity of $200\times$ the critical overdensity) are traced back
to their coordinates in the unperturbed initial particle distribution.
A contiguous surface enclosing the particles is identified out to
$3\times$ the virial radius, and the volume it encloses is resampled
with a high-resolution particle load realised by tiling a cubic
periodic glass distribution \citep{whi94}.  This high-resolution
particle load is divided into SPH and high-resolution dark matter
(DM1) particles, using $\Omega_{\rm b}=0.04825$ and $\Omega_{\rm
  m}-\Omega_{\rm b}=0.259$ respectively.  The remaining volume within
the simulation domain is resampled with an undivided load of
collisionless particles whose resolution decreases as a function of
distance from the surface ensuring the faithful reproduction of
large-scale gravitational forces throughout the high-resolution
region.  A glass-like particle distribution guarantees that the
initial particle load is essentially free of unwanted power above shot
noise frequency.  The high-frequency components of the Gaussian random
field perturbations can then faithfully be imposed on the zoom region.
The initial conditions are generated at $z=127$.

\subsubsection{Selected haloes}

We simulate two sets of haloes, a sample of ten $L^*$ haloes selected
from the EAGLE Recal-L025N0752 simulation, and a sample of ten
``group-sized'', super-$L^*$ haloes from the Ref-L100N1504 simulation.
The $L^*$ galaxy haloes are referred to as Gal001-Gal010 and the group
haloes are referred to as Grp000-Grp009 in Table 1.  We run at three
resolutions, which we will refer using the nomenclature {\it
  M}[log($m_{\rm SPH}/\msolar$)], where $m_{\rm SPH}$ is the initial
mass of SPH particles.

\begin{table*} \label{tab:zooms}
\caption{{\it M5.3} zoom simulation runs}
\begin{tabular}{ccrrrrrrrl}
\hline
Name$^a$ & 
Resolution$^b$ &
log $M_{200}^c$ ($\msolar$) &
$m_{\rm SPH}^d$ ($\msolar$) &  
$m_{\rm DM1}^d$ ($\msolar$) &
$\epsilon$ (pc) &
log $M_{*}^c$ ($\msolar$) &
SFR$^c$ ($\msolaryr$) &
$z_{\rm NEQ}^e$ & 
$z_{\rm low}^f$
\\
\hline
\multicolumn {8}{c}{}\\
Gal001  & {\it M5.3} & 12.07 & 2.22e+05 & 1.19e+06 & 350 & 10.27 & 1.426 & 0.503 & 0.0\\
Gal002  & {\it M5.3} & 12.25 & 2.28e+05 & 1.22e+06 & 350 & 10.26 & 1.275 & 0.503 & 0.0\\
Gal003  & {\it M5.3} & 12.11 & 2.32e+05 & 1.24e+06 & 350 & 10.25 & 1.717 & 0.503 & 0.0\\
Gal004  & {\it M5.3} & 11.99 & 2.26e+05 & 1.21e+06 & 350 & 10.08 & 0.881 & 0.503 & 0.0\\
Gal005  & {\it M5.3} & 12.17 & 2.20e+05 & 1.18e+06 & 350 & 10.40 & 2.919 & 0.503 & 0.0\\
Gal006  & {\it M5.3} & 11.91 & 2.22e+05 & 1.19e+06 & 350 & 10.08 & 1.021 & 0.503 & 0.0\\
Gal007  & {\it M5.3} & 11.82 & 2.17e+05 & 1.16e+06 & 350 &  9.93 & 1.695 & 0.503 & 0.0\\
Gal008  & {\it M5.3} & 11.85 & 2.27e+05 & 1.22e+06 & 350 &  9.95 & 1.030 & 0.503 & 0.0\\
Gal009  & {\it M5.3} & 11.85 & 2.27e+05 & 1.22e+06 & 350 &  9.94 & 1.117 & 0.503 & 0.0\\
Gal010  & {\it M5.3} & 12.67 & 2.21e+05 & 1.19e+06 & 350 & 10.67 & 0.183 & 0.503 & 0.0\\
Grp000  & {\it M5.3} & 12.74 & 2.27e+05 & 1.22e+06 & 350 & 10.70 & 0.130 & 0.282 & 0.0\\
Grp001  & {\it M5.3} & 12.75 & 2.01e+05 & 1.08e+06 & 350 & 10.65 & 3.036 & 0.282 & 0.205\\
Grp002  & {\it M5.3} & 12.77 & 2.41e+05 & 1.29e+06 & 350 & 10.79 & 2.065 & 0.282 & 0.0\\
Grp003  & {\it M5.3} & 12.73 & 2.38e+05 & 1.28e+06 & 350 & 10.76 & 2.113 & 0.282 & 0.0\\
Grp004  & {\it M5.3} & 12.89 & 2.38e+05 & 1.28e+06 & 350 & 10.74 & 0.373 & 0.282 & 0.0\\
Grp005  & {\it M5.3} & 12.98 & 2.34e+05 & 1.25e+06 & 350 & 10.75 & 0.021 & 0.282 & 0.149\\
Grp006  & {\it M5.3} & 13.01 & 2.22e+05 & 1.19e+06 & 350 & 10.57 & 0.017 & 0.282 & 0.0\\
Grp007  & {\it M5.3} & 12.88 & 2.38e+05 & 1.28e+06 & 350 & 10.60 & 5.574 & 0.503 & 0.0\\
Grp008  & {\it M5.3} & 13.19 & 2.31e+05 & 1.24e+06 & 350 & 10.95 & 0.063 & 0.282 & 0.205\\
Grp009  & {\it M5.3} & 13.19 & 2.31e+05 & 1.24e+06 & 350 & 10.90 & 1.188 & 0.282 & 0.0\\
\hline
\end{tabular}
\\
\parbox{25cm}{
  $^a$ ``Gal'' indicates $L^*$ haloes, ``Grp'' indicates group-sized haloes.\\ 
  $^b$ M[log$_{10}$($m_{\rm SPH}/\msolar$)]\\
  $^c$ At $z=0.205$\\
  $^d$ SPH and DM1 particle masses vary slightly between different zoom initial conditions.\\ 
  $^e$ $z_{\rm NEQ}$ is the redshift when the NEQ ionization and cooling are turned on.  \\
  $^f$ $z_{\rm low}$ is the redshift to which the NEQ simulations are run.
}
\end{table*}

\noindent{\bf {\it M5.3} Zooms:} We simulate all haloes at our
fiducial resolution, corresponding to an SPH particle mass resolution
of $2.3\times 10^5 \msolar$, and a dark matter (DM) particle mass of
$1.2\times 10^6 \msolar$.  The Plummer-equivalent softening length is
350 proper pc below $z=2.8$, and 1.33 comoving kpc above $z=2.8$,
which corresponds to 1/25th of the mean interparticle separation.
This is the same resolution as achieved by the L025N0752 simulation.
We can cross-reference the zoom $L^*$ {\it M5.3} galaxies with the
same galaxies in the L025N0752 galaxies, which we do in Appendix
\ref{sec:resconv} to ensure that the zooms do not differ substantially
from their parent cosmological volume.  We also cross-reference the
group zooms, although note that these zooms have $8\times$ ($2\times$)
higher mass (spatial) resolution than the L100N1504 simulation.  All
{\it M5.3} haloes have NEQ runs between the redshift $z_{\rm NEQ}$ and
$z_{\rm low}$ as explained in \S\ref{sec:runnoneq}, and each has also
been run using equilibrium cooling to $z=0$.  

\noindent{\bf {\it M4.4} Zooms:} We simulate the L025N0752 haloes at
$8\times$ higher resolution, corresponding to $m_{\rm SPH}=
2.7\times10^4 \msolar$ ($m_{\rm DM}=1.6 \times 10^5 \msolar$), and a
softening length of 175 proper pc below $z=2.8$, and 655 comoving pc
at earlier times.  All of these are run in equilibrium, and a
selection in non-equilibrium.  We only use these zooms to test
resolution convergence in Appendix \ref{sec:resconv} and they are
listed there in Table \ref{tab:zooms_app}.  These zooms confirm the
main CGM results reported throughout; however their stellar masses are
significantly lower, which is why we only use them for resolution
convergence tests.  

\noindent{\bf {\it M6.2} Zooms:} We simulate a selection of haloes at
the resolution of the L100N1504 box, $m_{\rm SPH}= 1.8\times 10^6
\msolar$ and $m_{\rm DM}= 9.7\times 10^6 \msolar$, and also compare
directly to the same galaxies in the 100 Mpc box, which allows us to
link our results to the global IGM results of \citet{rah15b}, who use
this box as their fiducial simulation.

Each zoom is focused on one halo, but surrounding simulated regions
often contain additional haloes that are either partially or fully
sampled by SPH and high-resolution DM1 particles.  We find a set of
``bonus'' haloes that we verified to be completely resolved with SPH
and DM1 particles out to three times the virial radius ($R_{200}$),
and add these to our sample.  Bonus haloes at our main redshift
($z=0.205$) include three sub-$L^*$ galaxies ($M_{200}\la 10^{11.5}
\msolar$) found at a distance greater than $3\times R_{200}$ of the
targeted $L^*$ haloes, two low-mass $L^*$ haloes ($M_{200}=10^{11.6}
\msolar$) and two $L^*$ haloes ($M_{200}=10^{12.0-12.5} \msolar$)
found beyond $3\times R_{200}$ of targeted group haloes, and three
bonus haloes ($M_{200}= 10^{12.0-12.6}$) in the Gal010 zoom, which
itself is by far the most massive ``Galaxy'' zoom, and tracks the
formation of a $10^{12.7} \msolar$ $z=0$ halo.

\subsubsection{Running in equilibrium}\label{sec:runequil}

We evolve all zooms using the equilibrium cooling rates from
\citet{wie09a} to $z=0$ using the standard EAGLE code with the only
difference being that we do not use kernel-smoothed metallicities to
compute cooling rates.  As explained by \citet{wie09b}, smoothed
metallicities are consistent with the smoothed particle hydrodynamic
formalism; however, their use when adopting NEQ cooling requires that
both particle and smoothed ion abundances are tracked, which for 136
species is a prohibitive memory requirement.  Furthermore, smoothing
complicates the interpretation of NEQ ionization effects when ion
species are averaged over the SPH kernel.  Therefore, to enable a
consistent comparison of methods, we use particle metallicities in
both equilibrium and NEQ runs.

As a consistency test, we run a set of five {\it M5.3} zoom runs with
smoothed metallicities, finding that they have $0.11$ dex higher
stellar masses than their non-smoothed {\it M5.3} zoom counterparts.
It appears that smoothed metallicities increase the SF efficiencies by
distributing metals over more particles, increasing the overall
cooling rate of the galaxy, and leading to more star formation.
\citet{wie09b} also showed that in the OWLS simulations, smoothed
metallicities increased the SF efficiency.  We show in Appendix
\ref{sec:resconv} that $\OVI$ column densities are unaffected by metal
smoothing.  

\subsubsection{Running in non-equilibrium} \label{sec:runnoneq}

When examining NEQ simulations, we activate the NEQ module at
relatively low redshift in order to minimize the computational cost of
following 136 ionic species.  The magnitude of non-equilibrium effects
can be significant in shocks and rapidly cooling gas, and this
behaviour is only captured with use of the NEQ module.  This module is
important to use, in order to recover the correct ion species in all
circumstances.  It should be noted that we will show that the NEQ
effects are generally not important for gas dynamics in a separate
study, as predicted by OS13 and also demonstrated by \citet{ric16}.
Furthermore, NEQ effects on CGM ion diagnostics (i.e. observables)
persist for short times compared to the Hubble timescale, enabling us
to begin modelling them only at late times.  We discuss the NEQ
effects in Appendix \ref{sec:NEQ}, which are typically small, $<0.1$
dex for oxygen ions.  

The $L^*$ haloes all switch to NEQ cooling at $z_{\rm NEQ}=0.503$ and
the group haloes at $z_{\rm NEQ}=0.282$.  The later NEQ activation for
group haloes was adopted to save computational cost.  The NEQ runs
begin from a snapshot output of the equilibrium run, where we iterate
{\tt CVODE} for a long timescale ($\delta t= 10^6 /$[$\nh/\cmc$] yrs)
for all species to guarantee ionization equilibrium.  The EAGLE code
then runs as normal with the \citet{wie09a} cooling replaced by the
NEQ module iterating the ionization states and cooling rates via {\tt
  CVODE} across the hydrodynamics timestep.

For one group zoom (Grp007) we perform two runs, $z_{\rm NEQ}=0.503$
and $0.282$ in order to establish that its properties are not
significantly altered at $z=0.250$, since this is the highest redshift
we use to compare to COS-Halos results.  We use up to six outputs
($z=0.250, 0.205, 0.149, 0.099, 0.053, 0.0$) when confronting
COS-Halos measurements to give a greater range of galaxy properties as
discussed further in \S\ref{sec:COS-Halos}.  Every halo is run past
$z=0.205$, with every {\it M5.3} $L^*$ halo run to $z=0$ and 7 out of
10 {\it M5.3} group haloes run to $z=0$.

\subsubsection{Basic galaxy properties}

\begin{figure*}
\includegraphics[width=0.49\textwidth]{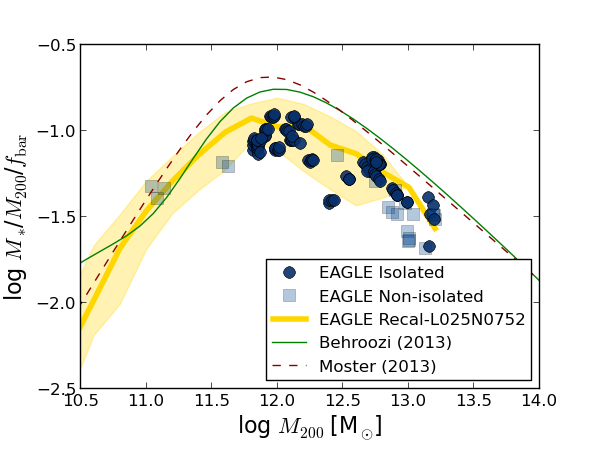}
\includegraphics[width=0.49\textwidth]{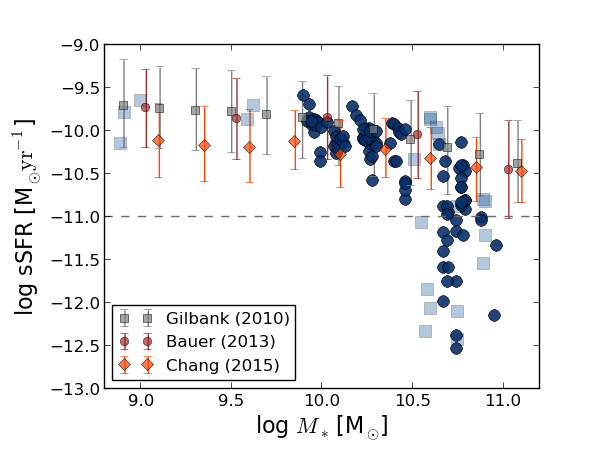}
\includegraphics[width=0.235\textwidth]{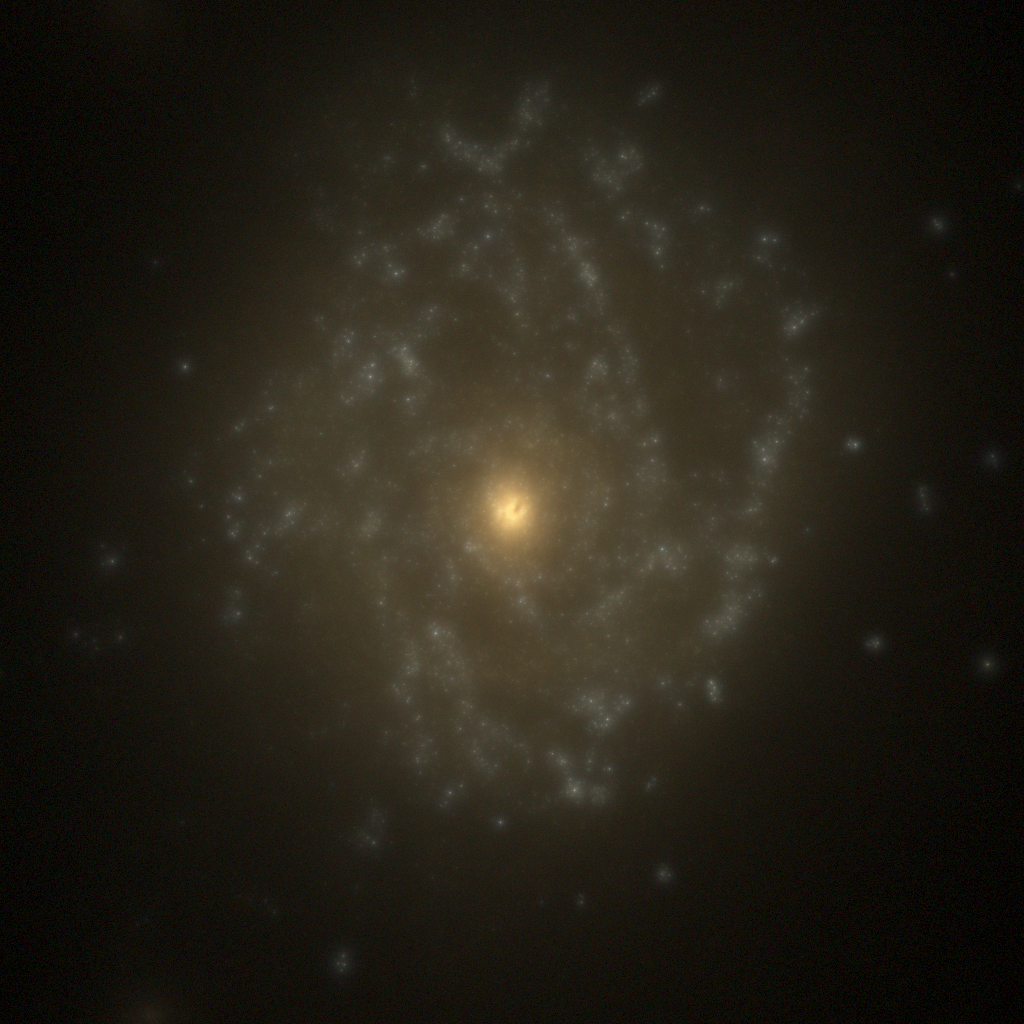}
\includegraphics[width=0.235\textwidth]{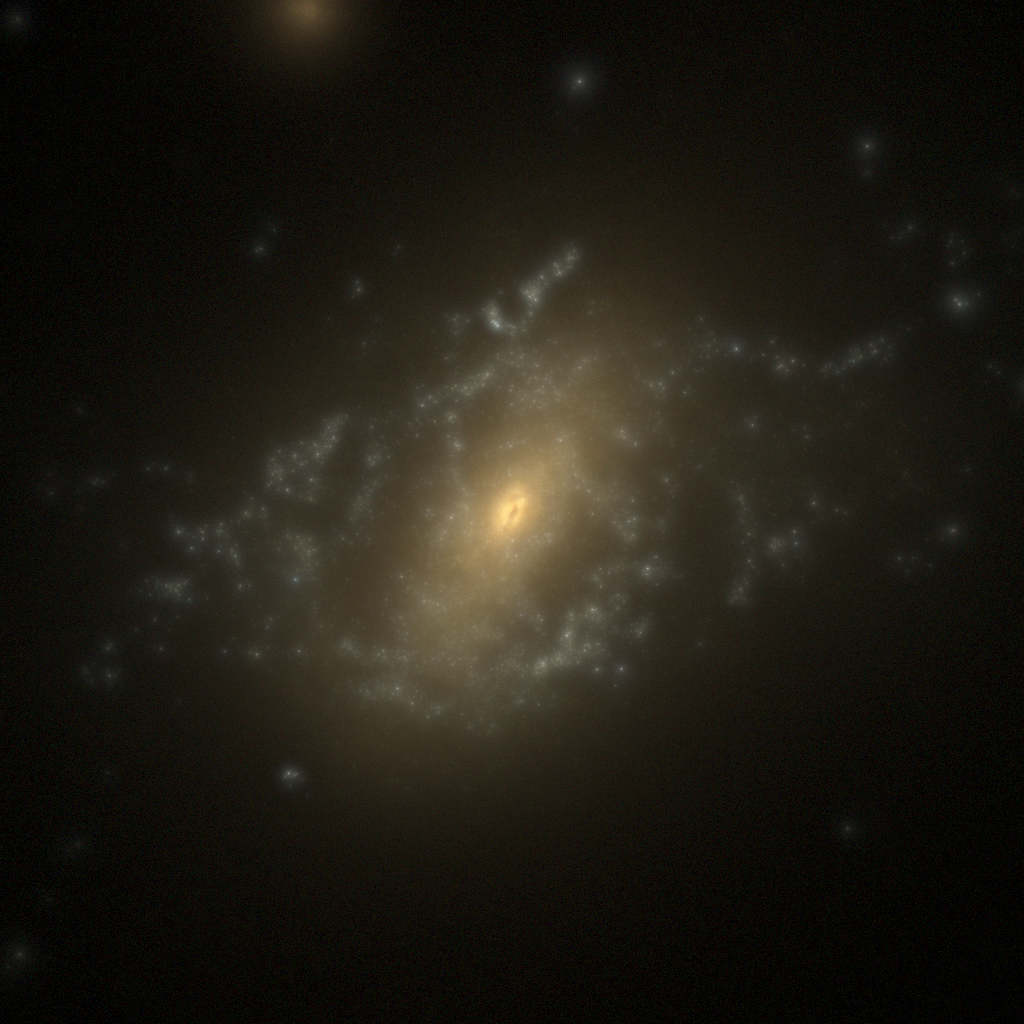}
\includegraphics[width=0.235\textwidth]{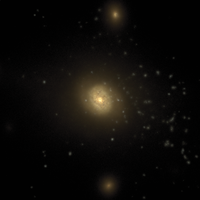}
\includegraphics[width=0.235\textwidth]{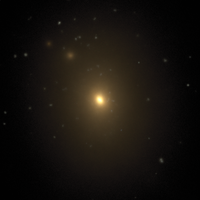}
\caption[]{Stellar mass to halo mass ratio plotted as a function of
  halo mass of zoomed centrals (top left) and sSFR as a function of
  stellar mass (top right) from 6 redshifts ($z=0.25\rightarrow 0$)
  that we use in our analysis.  Galaxies that can be considered
  ``isolated'' by COS-Halos criteria are indicated by dark circles,
  while transparent squares indicate galaxies with a neighboring
  galaxy with $M_*>10^{10.3} \msolar$ within 300 kpc.  Three sub-$L^*$
  galaxies at $z=0.205$ are also plotted as light circles.  The thick
  yellow line is the fit to all centrals in the Recal-L025N0752
  simulation (S15) with the shading indicating the 1-$\sigma$
  dispersion.  Abundance matching results from \citet[][solid
    green]{beh13a} and \citet[][dashed red]{mos13} at $z=0.2$ are
  displayed on the left, and observations of the sSFR of star-forming
  galaxies as a function of $M_{*}$ from \citet[][grey
    squares]{gil10}, \citet[][brown circles]{bau13}, and
  \citet[][orange diamonds]{cha15} are displayed on the right.  Bottom
  panels show simulated 50 kpc SDSS $u$, $g$, and $r$-band composite
  images, including dust, of two star-forming $L^*$ galaxies and two
  group centrals-- a recent merger and a passive galaxy (from left to
  right).  All plots use {\it M5.3} zooms.}
\label{fig:abundance_ssfr}
\end{figure*}

We plot stellar mass and SFR properties of central galaxies in our
haloes in Figure \ref{fig:abundance_ssfr} using {\it M5.3} zooms.  We
run the {\tt SUBFIND} algorithm \citep{spr01, dol09} on our zoom
outputs, and quantify halo mass, $M_{200}$, as the mass within a
sphere with mean enclosed density $200\times$ the critical overdensity
centred on the galaxy's potential minimum.  Galaxy stellar masses
($M_*$) and star formation rates (SFRs) are calculated by summing the
appropriate quantities within a 30 physical kpc sphere around the
centre of the central subhalo, which almost always hosts the most
massive galaxy.  The top left panel of Fig. \ref{fig:abundance_ssfr}
plots the $M_*/(M_{200}\times f_{\rm bar})$ abundance matching
relation, where $f_{\rm bar}\equiv\Omega_{\rm b}/\Omega_{\rm m}$, for
central galaxies from the six redshifts we examine.  While our {\it
  M5.3} haloes follow the shape of the relationship recovered by
\citet{beh13a} and \citet{mos13} at $z=0.2$, they are $\sim 0.2-0.4$
dex lower for $M_{200}\ga 10^{12}\msolar$.  As discussed earlier, our
use of particle metallicities reduces stellar masses by 0.1 dex
compared with the smoothed metallicity simulations used to calibrate
the feedback efficiencies.  Moreover, S15 also find stellar masses
0.2-0.3 dex lower than the abundance matching constraints
\citep{beh13a, mos13} at $M_{200} = 10^{12.0} \msolar$ at $z=0.1$
(Recal-L025N0752 shown as a thick yellow line with shading for
1-$\sigma$ dispersion).

For each central galaxy at a given redshift, we test whether it would
be classified as isolated using criteria similar to those applied by
\citet{tum13}, who select COS-Halos galaxies as being the most
luminous galaxy within $b<300$ kpc of the QSO sightline at its
redshift.  They also note that there are usually no $L^*$ photo-$z$
candidates within 1 Mpc.  We use the criterion that there should not
be any galaxies with $M_*>2\times 10^{10} \msolar$ within an impact
parameter $b=300$ kpc, firstly to ensure that our galaxy is the most
massive system inside this impact parameter, and secondly because we
usually do not simulate surrounding galaxies beyond $3 R_{200}$.
Often there are surrounding haloes that are not resolved completely by
the high-resolution particle load; however we still use their stellar
masses in our isolation criteria to be conservative, although there
are only a few cases for which this happens.  Dark circles in
Fig. \ref{fig:abundance_ssfr} indicate haloes that appear isolated in
at least one of the three projections we use to cast our sight lines
and can be used for comparison with the COS-Halos dataset (see
\S\ref{sec:COS-Halos}).  Transparent squares are for non-isolated
haloes in all three projections, as well as for bonus haloes with
$M_{200}<10^{11.7} \msolar$, where we only plot their outputs at
$z=0.205$.  We include these low-mass haloes for our discussion of
halo mass trends in \S\ref{sec:halotrends}, but not for comparisons to
the COS-Halos dataset in \S\ref{sec:COS-Halos}.

We plot the galaxies on the $M_*$-sSFR plot in the top right panel of
Fig. \ref{fig:abundance_ssfr} to show that our galaxies follow the
general trend of the star-forming sequence observed at low-$z$ by
\citet{gil10}, \citet{bau13}, and \citet{cha15}, and a set of massive
galaxies in group haloes with low sSFR.  Our galaxies do not have a
passive clump at sSFR$<10^{-11}$ yr$^{-1}$ that is as prominent as for
the COS-Halos survey, but the EAGLE boxes do reproduce the passive
fraction as a function of $M_*$ (S15).  Additionally, when we mock
COS-Halos surveys in \S\ref{sec:COS-Halos}, the galaxies we select to
match the COS-Halos galaxies better reproduce this passive clump.  The
bottom row of Fig. \ref{fig:abundance_ssfr} shows 50 kpc, SDSS $u$,
$g$, and $r$-band composite maps of two star-forming $L^*$ galaxies
(left), and two group centrals, including an active merger and a
passive galaxy (right).  These maps are generated using the SKIRT
radiative transfer code \citep{bae11, cam15} as described in detail by
Trayford et al. (in prep.).

\section{Physical properties of the oxygen-enriched circumgalactic medium} \label{sec:phys}

This section begins by exploring CGM properties across two decades in
halo mass, from sub-$L^*$ haloes ($M_{200}\la 10^{11.7} \msolar$)
through $L^*$ haloes ($M_{200}\sim 10^{12} \msolar$) to group-sized,
super-$L^*$ haloes ($M_{200}\ga 10^{12.3} \msolar$).  We further
explore the physical trends that establish the ionization structure of
the oxygen enriched CGM.  Lastly, we focus on the nature and origin of
the $\OVI$ that is observed by the COS-Halos survey.  Throughout we use
{\it M5.3} zooms.

\subsection{Trends with halo mass} \label{sec:halotrends}

\begin{figure*}
\includegraphics[width=0.32\textwidth]{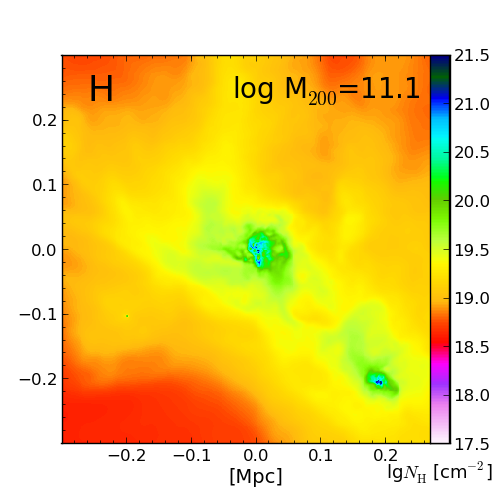}
\includegraphics[width=0.32\textwidth]{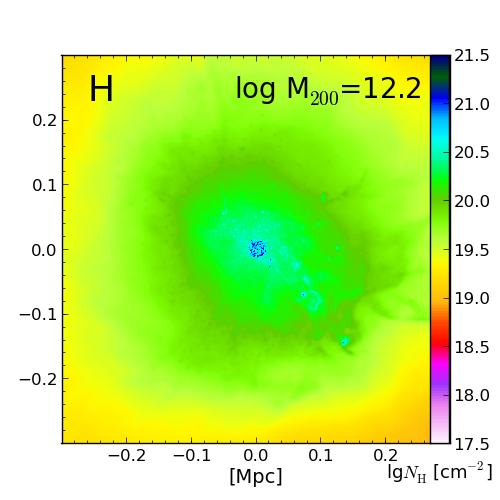}
\includegraphics[width=0.32\textwidth]{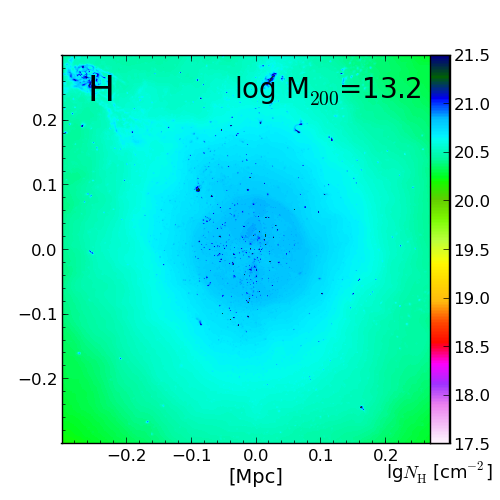}
\includegraphics[width=0.32\textwidth]{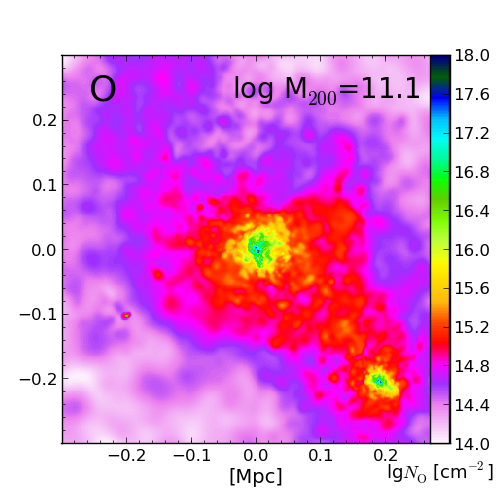}
\includegraphics[width=0.32\textwidth]{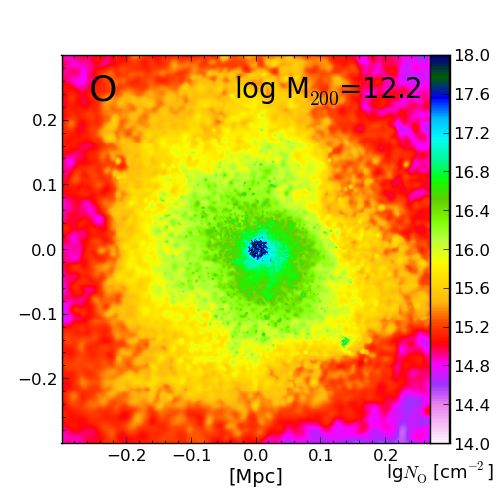}
\includegraphics[width=0.32\textwidth]{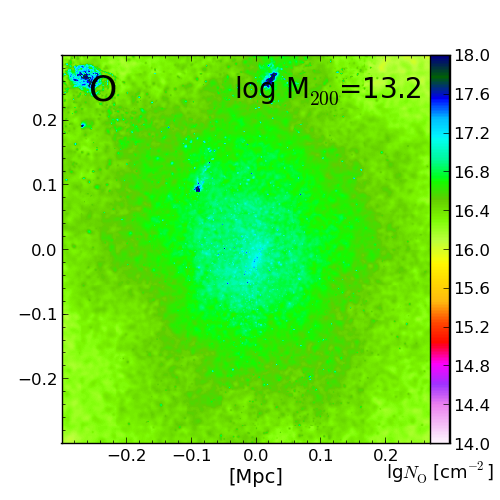}
\includegraphics[width=0.32\textwidth]{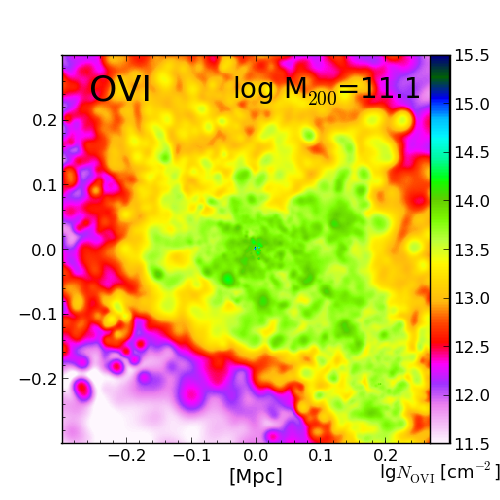}
\includegraphics[width=0.32\textwidth]{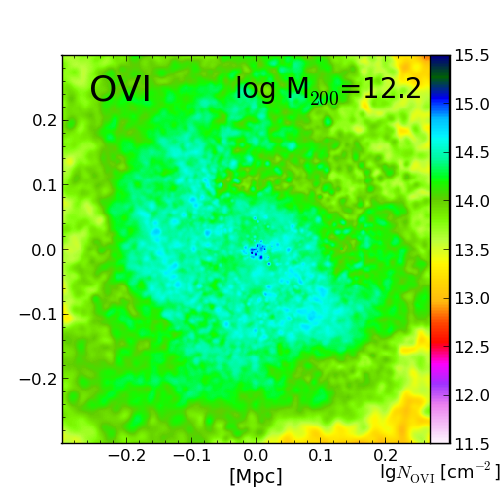}
\includegraphics[width=0.32\textwidth]{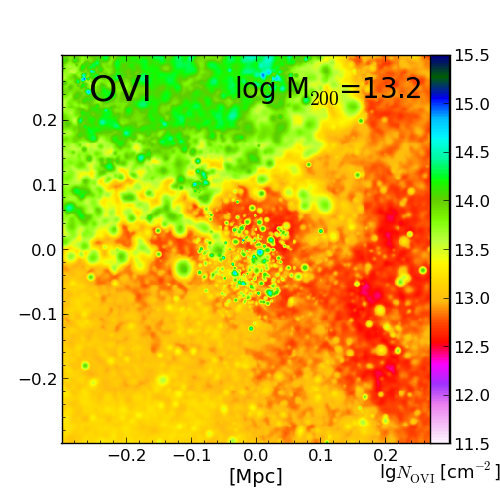}
\includegraphics[width=0.32\textwidth]{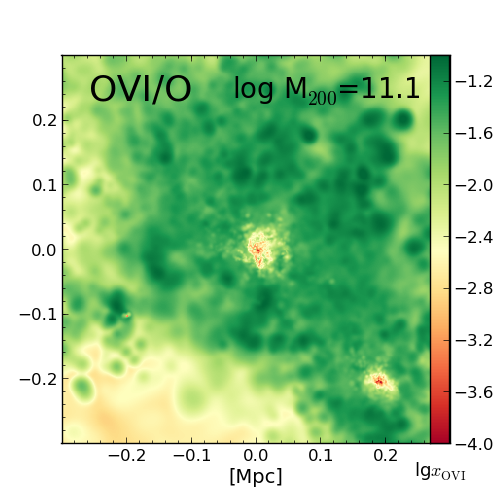}
\includegraphics[width=0.32\textwidth]{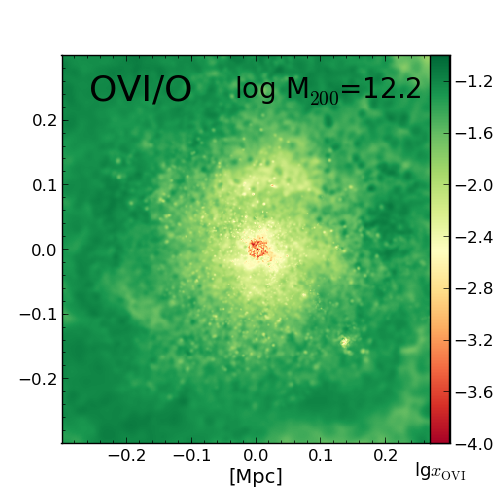}
\includegraphics[width=0.32\textwidth]{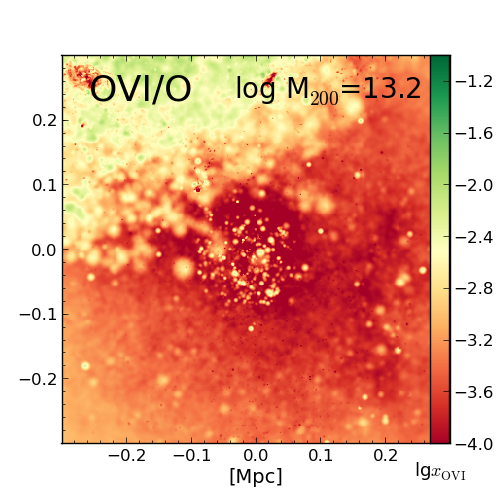}
\caption[]{Images of 600 physical kpc on a side $z=0.205$ snapshots of
  three haloes with mass $10^{11.1}, 10^{12.2}, \& 10^{13.2} \msolar$
  representative of sub-$L^*$, $L^*$, and group-sized haloes,
  respectively, from left to right.  The top row shows hydrogen
  columns.  The 2nd row shows oxygen columns, while the 3rd row shows
  $\OVI$ columns.  The bottom row shows maps of the $\OVI$ fraction.
  The selected galaxies are a bonus sub-$L^*$ central galaxy in the
  Gal005 zoom, the main $L^*$ galaxy in the Gal005 zoom, and the
  super-$L^*$ galaxy in the Grp008 zoom.  Projections are 2400 kpc
  deep.}
\label{fig:halotrends}
\end{figure*}

We begin our discussion by showing three haloes with masses
$M_{200}=10^{11.1}$, $10^{12.2}$, and $10^{13.2} \msolar$ in Figure
\ref{fig:halotrends} at the median redshift of the COS-Halos, $z=0.2$.
All images are 600 physical kpc across.  The central galaxies have
$M_*= 10^{9.0}$, $10^{10.4}$, and $10^{11.0} \msolar$, and typical
SFRs for their masses.

\begin{figure*}
\includegraphics[width=0.32\textwidth]{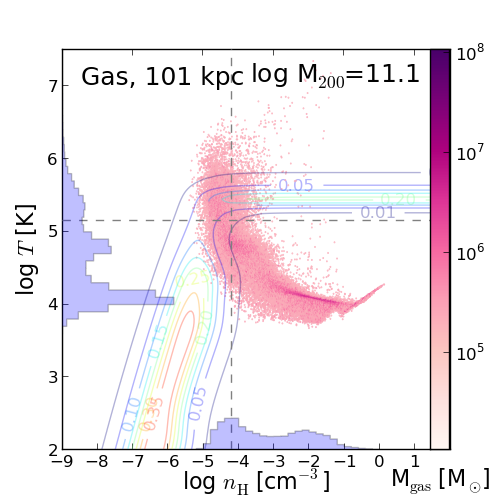}
\includegraphics[width=0.32\textwidth]{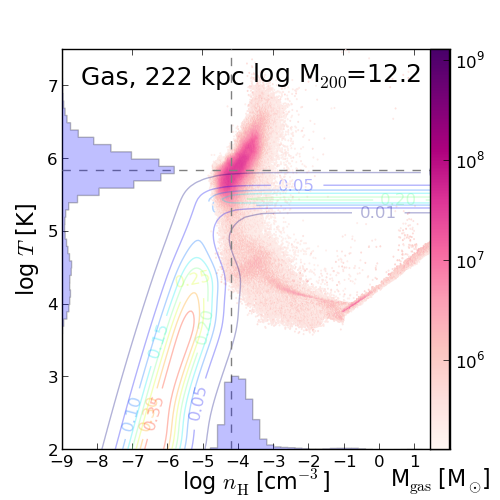}
\includegraphics[width=0.32\textwidth]{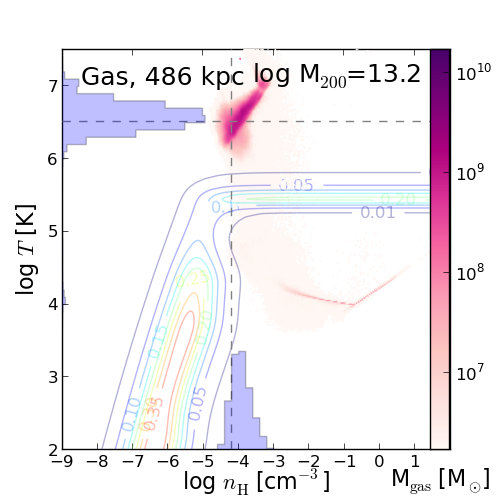}
\includegraphics[width=0.32\textwidth]{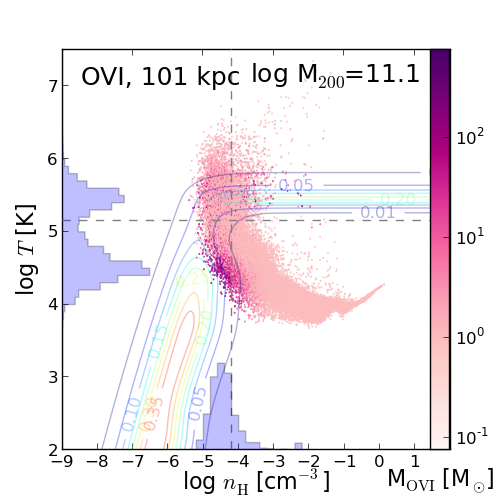}
\includegraphics[width=0.32\textwidth]{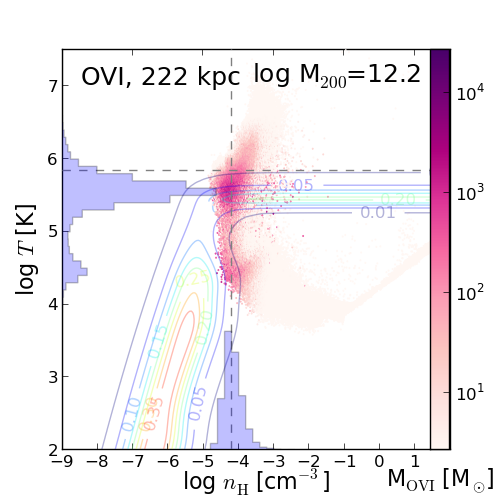}
\includegraphics[width=0.32\textwidth]{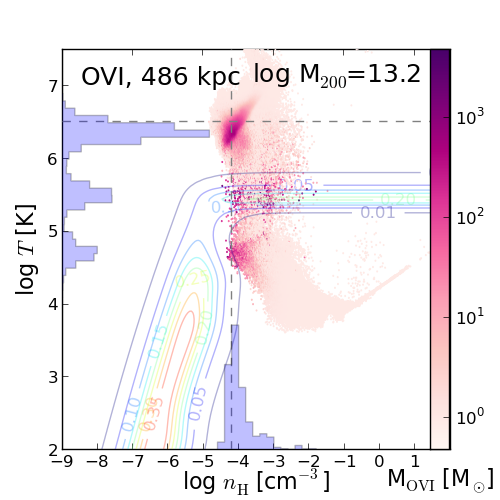}
\caption[]{The mass distribution in the density-temperature phase
  space of the haloes shown in Fig. \ref{fig:halotrends} at $z=0.205$.
  The top row shows all gas within $R_{200}$ (listed in the top left
  in kpc) and the bottom row shows only $\OVI$.  Histograms in each
  panel represent the probability density functions of gas density
  (bottom) and temperature (left).  The dashed grey horizontal line
  indicates $T_{\rm vir}$ (Equ. \ref{equ:Tvir}), and the dashed grey
  vertical line indicates $200\times$ the critical overdensity.
  Contours show $\OVI$ ionization fraction levels with collisional
  ionization contours peaking at $10^{5.5}$ K, and photo-ionization
  contours at $<10^5$ K for lower densities.}

\label{fig:halotrends_rhot_Rvir}
\end{figure*}

We show hydrogen column densities in the top row with halo mass
increasing from left-to-right.  We plot the $\rho-T$ diagnostic
diagrams of all gas inside of $R_{200}$ in Figure
\ref{fig:halotrends_rhot_Rvir} (top row), and see that the temperature
histograms peak at $T_{\rm peak}=10^{4.1}$, $10^{5.8}$, and $10^{6.6}$
K.  We argue in \S\ref{sec:oxycont} that the typical temperature of
halo gas roughly follows that given by an isothermal, single-phase
virialized halo,

\begin{equation} \label{equ:Tvir}
 T_{\rm vir}= 10^{5.69} {\rm K} \left(\frac{M_{200}}{10^{12} \msolar} \right)^{2/3} \left(\Omega_{\rm M}(1+z)^3+\Omega_{\Lambda}\right)^{2/3}
\end{equation}

\noindent using the virial theorem.  Using Equ. \ref{equ:Tvir}, we
recover $T_{\rm vir}=10^{5.1}$, $10^{5.8}$, and $10^{6.5}$ K
(indicated by dashed grey horizontal lines in
Fig. \ref{fig:halotrends_rhot_Rvir}).  For the sub-$L^*$ halo, the
$T_{\rm peak}\ll T_{\rm vir}$ owes to efficient cooling through the
peak of the cooling curve to thermal equilibrium at $\sim 10^4$ K,
although there does exist a tertiary peak near $T_{\rm vir}$.  The
more massive haloes have peak temperatures more similar to $T_{\rm
  vir}$.

The virial radius across these haloes increases by nearly a factor of
five from 101 to 486 kpc, meaning that the adoption of a constant
radius for the CGM, often assumed to be 300 kpc
\citep[e.g.][]{rud12,for13}, samples only the inner region of a group
halo, but includes the diffuse IGM for a sub-$L^*$ halo.  It is
important to remember that the ratio of the impact parameter to the
virial radius varies significantly across a survey such as COS-Halos
\citep{shu14}.

The second row of Fig. \ref{fig:halotrends} shows that oxygen columns
increase with halo mass, and our overall sample indicates that there
is an increase of $N_{\Oxy}$ with $M_*$ at all impact parameters,
except at $b<30$ kpc where $L^*$ galaxies have higher oxygen columns
than groups.  The mass of gaseous oxygen within a 300 kpc sphere
increases by a factor of 40 as $M_{200}$ increases by a factor of 100,
with the greatest increase occurring from sub-$L^*$ to $L^*$
($17\times$), while $L^*$ galaxies have the highest metallicities
within 300 kpc ($Z/Z_{\odot}=0.11$, $0.36$, $0.22$; excluding
star-forming gas does little to change these values).

The third row of Fig. \ref{fig:halotrends} shows the column densities
of $\OVI$.  It is striking that the $L^*$ galaxies have by far the
greatest $\OVI$ column density even though they do not have the
highest gas and oxygen masses. The mass of $\OVI$ within a 300 kpc
sphere increases by a factor of 9 from sub-$L^*$ to $L^*$, but
declines by a factor of 90 from the $L^*$ to the group halo.  

The bottom row divides the $\OVI$ map by the oxygen map to show the
``ionization fraction,'' which severely declines for the group halo.
The global $\OVI$ fraction within a 300 kpc sphere falls from
$x_{\OVI}=2.8\%$ to $1.4\%$ from sub-$L^*$ to $L^*$, and to $0.01\%$
for the group halo.  $\OVI$ is merely the tip of the iceberg of the
oxygen content in the CGM of star-forming galaxies.  Meanwhile, the
oxygen in group-sized haloes essentially disappears in the ``$\OVI$
filter''.  The higher $\OVI$ fractions on the top left of the group
halo (bottom right panel) owes to a lower mass, foreground halo with
higher CGM $\OVI$.  

The $\rho-T$ diagrams of gas inside $R_{200}$ in Fig.
\ref{fig:halotrends_rhot_Rvir} clearly highlight the cause of this
effect: groups exhibit virtually no gas below $2\times10^6$ K (upper
right panel).  Regions with gas overlapped by $\OVI$ ionization
fraction contours indicate where $\OVI$ can exist in significant
quantities, and the $\OVI$-traced gas is shown in the bottom panels.
The triple-peak temperature histogram of the $\OVI$ distribution in
the group (y-axis histogram, bottom right panel) is common among our
group haloes: metals at $T>10^6$ K and $T<10^5$ K are abundant due to
longer cooling times but have $x_{\OVI}<1\%$, while a small number of
particles are rapidly cooling through the coronal regime ($10^5-10^6$
K) and briefly achieve $x_{\OVI}\sim 20\%$.  In contrast, $L^*$ halo
virial temperatures, $T_{\rm vir} \sim 0.5-1 \times 10^6$ K, overlap
with the collisionally ionized $\OVI$ band at $3\times 10^5$ K (bottom
centre panel).

\begin{figure*}
\includegraphics[width=0.49\textwidth]{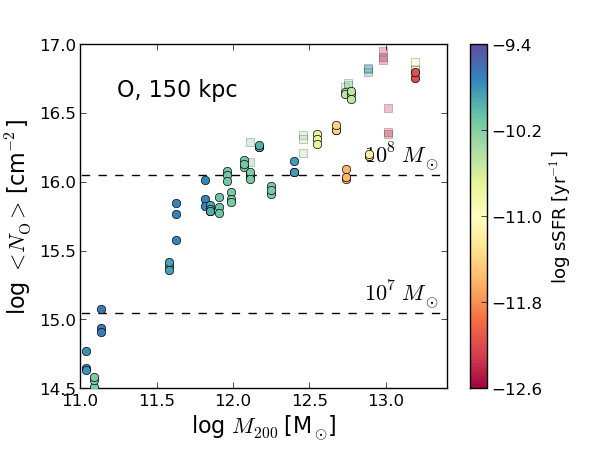}
\includegraphics[width=0.49\textwidth]{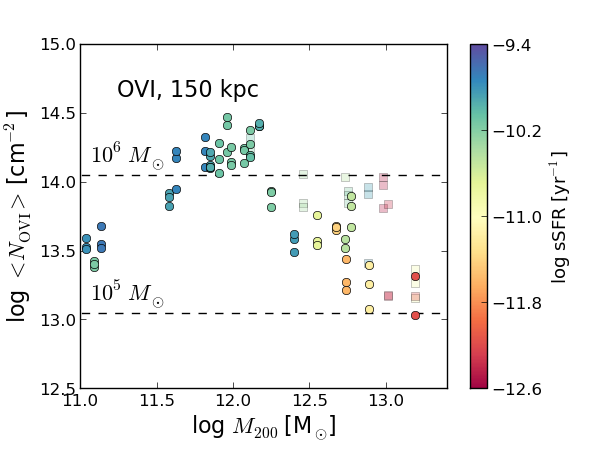}
\caption[]{Aperture column densities within $b=150$ kpc for oxygen
  (left) and $\OVI$ (right) as a function of halo mass at $z=0.205$.
  Equivalent summed masses within 150 kpc are marked by dashed
  lines. Data points are coloured by sSFR, according to the colourbar
  on the right.  Up to three data points for each halo are for three
  projections, with solid circles showing isolated projections and
  transparent squares that are lightly shaded indicating projections
  that do not satisfy the isolation criteria.}
\label{fig:mhalo_NOVI}
\end{figure*}

The sub-$L^*$ CGM with gas between $10^4-\ga 10^5$ K overlaps the
photo-ionized $\OVI$ contours, which appear in gas $\la 10^5$ K at
lower densities \citep[e.g.][]{shu12}.  $\OVI$ peaks at a lower
density ($\nh=10^{-4.5} \cmc$) in the photo-ionized regions of the
sub-$L^*$ halo than in the collisionally ionized regions of the $L^*$
halo ($10^{-4.2} \cmc$). \citet{opp09a} showed that most $\OVI$
absorbers are photo-ionized using other SPH simulations, because the
\citet{haa01} background is strong enough to photo-ionize IGM oxygen,
which dominates the $\OVI$ absorbers found along intervening sight
lines.  The higher densities combined with hotter gas means that
collisionally ionized $\OVI$ traces $L^*$ CGM gas.

$L^*$ haloes also have higher $N_{\OVI}$ (Fig. \ref{fig:halotrends}).
To demonstrate that this is a broad trend across all of our haloes, we
plot 150 kpc ``aperture'' columns for oxygen (left) and $\OVI$ (right)
in Figure \ref{fig:mhalo_NOVI} for all galaxies, where we calculate
the aperture column within a cylinder of radius $b=150$ kpc by summing
up pixel column densities, $N(x,y)$, as plotted in
Fig. \ref{fig:halotrends} according to

\begin{equation} \label{equ:apcol} 
\langle N \rangle_{b} =  \frac{\sum\limits_{<b} N(x,y) dx^2}{\pi b^2} \cms
\end{equation}

\noindent where $dx$ is the pixel size such that $dx\ll b$.  We take
aperture columns for each halo in three projections (along the $x$,
$y$, \& $z$ axis), and show isolated haloes as solid circles and
non-isolated counterparts as transparent squares.  $\langle
N_{\Oxy}\rangle_{150}$ increases monotonically across our halo mass
range, while $\langle N_{\OVI}\rangle_{150} $ peaks for $L^*$ galaxies
and then declines sharply for $M_{200}>10^{12.3} \msolar$.  The
average $\langle N_{\OVI}\rangle_{150} $ for the $L^*$ range
($M_{200}= 10^{11.7}-10^{12.3} \msolar$) is $10^{14.23} \cms$
corresponding to an integrated $\OVI$ mass of $1.5\times 10^6 \msolar$
within a cylinder of radius 150 kpc.  The corresponding values for
groups ($M_{200}= 10^{12.3}-10^{13.3} \msolar$) are $10^{13.50} \cms$
and $2.8 \times 10^5 \msolar$, or five times lower.  We use only
isolated galaxies in these calculations, but by including non-isolated
projections, the $\langle N_{\OVI}\rangle_{150}$ in group-sized haloes
rises to $10^{13.69} \cms$, while that of $L^*$ haloes does not
change.  $\langle N_{\Oxy}\rangle_{150}$ equals $10^{15.99} \cms$ for
$L^*$ haloes and $10^{16.51} \cms$ for group haloes using isolated
samples.

\begin{figure*}
\includegraphics[width=0.49\textwidth]{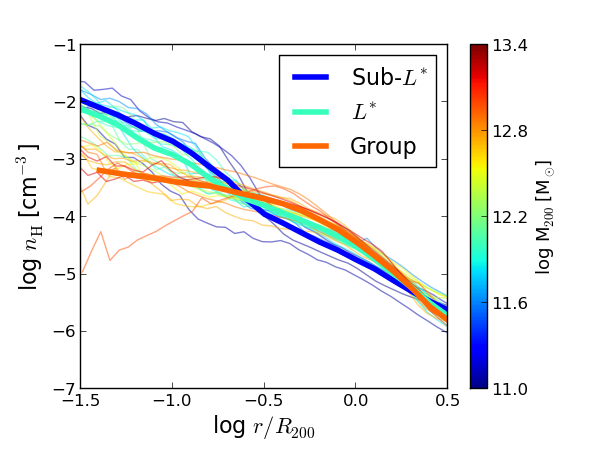}
\includegraphics[width=0.49\textwidth]{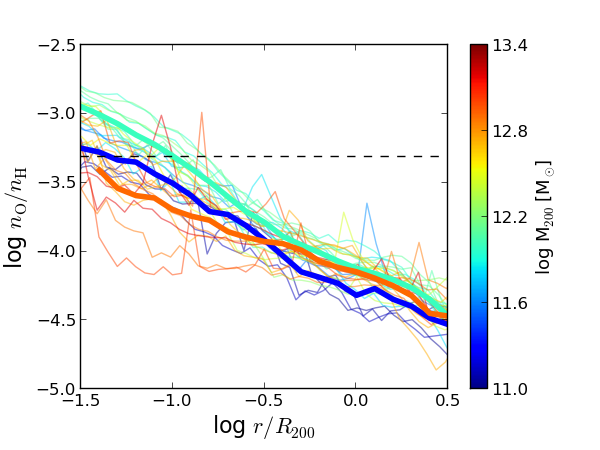}
\includegraphics[width=0.49\textwidth]{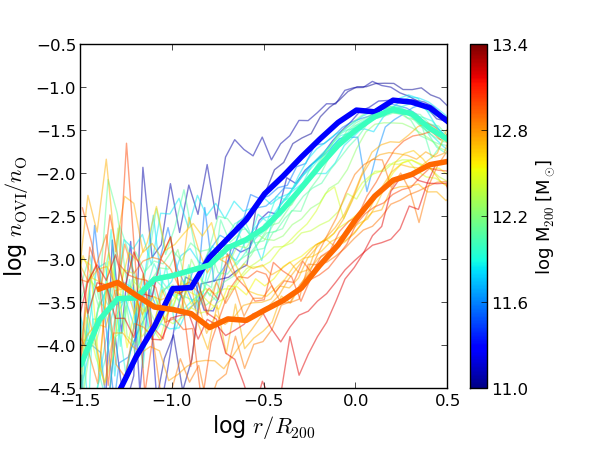}
\includegraphics[width=0.49\textwidth]{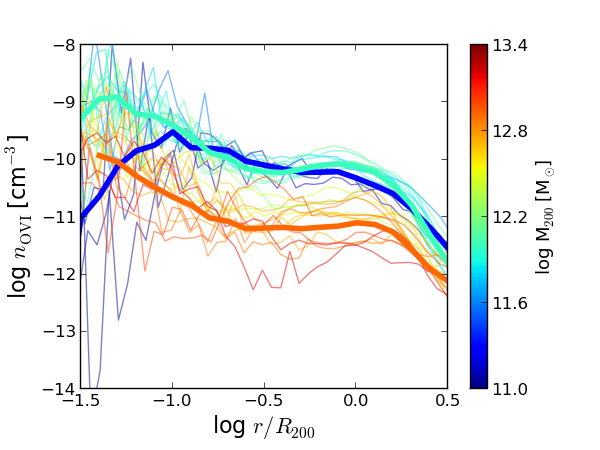}
\caption[]{Radial profiles averaged in spherical shells of the
  hydrogen density ($\nh$), oxygen abundance ($n_{\rm O}/\nh$), $\OVI$
  ion fraction ($x_{\OVI}\equiv n_{\OVI}/n_{\rm O}$), and $\OVI$
  density ($n_{\OVI}$) as a function of fractional virial radius from
  top left to bottom right at $z=0.205$.  Individual haloes are shown
  as thin lines, and thick solid lines represent averages of
  sub-$L^*$, $L^*$, and group-sized haloes (blue, aquamarine, \&
  orange).  The dashed line in the top right panel indicates the solar
  oxygen abundance ($n_{\rm O}/\nh=10^{-3.31}$).  Lower mass haloes
  have higher densities in their centres (top left), and $L^*$ mass
  haloes have the highest oxygen abundances in their centres (top
  right).  $x_{\OVI}$ (bottom left) shows a strong trend with halo
  mass.  Sub-$L^*$ haloes ($<10^{11.5} \msolar$) have the highest
  $x_{\OVI}$ owing to higher photo-ionized fractions, but this
  photo-ionized phase is greatly reduced in $L^*$ haloes owing to
  higher pressures.  Instead, $L^*$ haloes are dominated by
  collisionally ionized $\OVI$, and ion fractions only reach $\sim
  5\%$ at radii just beyond $R_{200}$.  Group haloes have even lower
  $x_{\OVI}$ because the vast majority of the gas is too hot for
  $\OVI$, and only gas beyond $R_{200}$ approaches $x_{\OVI}$ of
  several percent.  The bottom right panel, showing $\OVI$ densities,
  is the product of the previous three panels, and shows the higher
  $\OVI$ densities for sub-$L^*$ and $L^*$ haloes than for group
  haloes.  }
\label{fig:rvir_OVI}
\end{figure*}

\subsection{Circumgalactic radial profiles} 

To better understand the halo trends of the CGM distributions with
mass, we show in Figure \ref{fig:rvir_OVI} physical gas properties as
a function of fractional virial radius at $z=0.205$.  The top left
panel shows $\nh$ including all gas (CGM+ISM) as a function of
$r/R_{200}$.  We use thick lines to indicate the mean profiles for
sub-$L^*$ ($M_{200}<10^{11.7} \msolar$, blue), $L^*$ ($10^{11.7}\leq
M_{200}<10^{12.3} \msolar$, aquamarine), and group-sized ($M_{200}\geq
10^{12.3}$, orange) haloes.  Sub-$L^*$ galaxies achieve the highest
inner densities and the lowest outer densities, while for groups the
inner radial profiles ($\la 1/3 R_{200}$) are nearly flat.  ISM gas,
defined as gas having SFR$>0$, is not included in these plots, so
there is more dense CGM gas around the lower mass galaxies that have
higher sSFR.  However, the density of extended ``coronal'' gas ($1/3
R_{200} \la r \la R_{200}$) is higher for group haloes, owing to
inefficient cooling at $>10^6$ K enabling them to retain higher
pressure gas.  All haloes are centred on the minimum of the
gravitational potential, although one $10^{12.9} \msolar$ halo is
undergoing a major merger, causing the density to dive inside $1/3
R_{200}$ and neither galaxy coincides precisely with the potential
minimum.

The gas metallicities in the top right panel are highest for $L^*$
haloes in the centre, lower for sub-$L^*$ throughout the entire halo,
and lower for group haloes in the inner regions.  The curves for
individual haloes are noisier in this plot compared to the density
plot, because satellite galaxies influence radial metallicities more
than radial densities; groups have the greatest number of satellites
and thus more peaks overall.  The bottom left panel showing the radial
$\OVI$ fraction ($x_{\OVI}$) shows as strong of a trend of any of the
panels in Fig. \ref{fig:rvir_OVI}.  The peak in $x_{\OVI}$ moves to
larger fractional radius and declines with increasing halo mass.  The
sub-$L^*$ galaxies achieve $x_{\OVI}$ as high as 10\% at $\sim
R_{200}$, because the lower pressures allow oxygen at photo-ionized
$\OVI$ temperatures, and therefore greater $\OVI$ fractions
\citep{opp09a}.  The $L^*$ haloes have little scatter in $x_{\OVI}$,
only reach $x_{\OVI}\sim3\%$ at $R_{200}$, and peak just beyond
$R_{200}$.  This is almost all collisionally ionized $\OVI$, and the
maximal radial $x_{\OVI}$ is $\sim5\%$, with only a fraction of oxygen
residing around $10^{5.5}$ K.  At $M_{200}>10^{12.3} \msolar$,
$x_{\OVI}$ essentially nosedives, owing to hotter temperatures.
Collisionally ionized $\OVI$ does exist at $\sim 10^{5.5}$ K in group
haloes, but the higher pressures mean higher densities, leading to
faster cooling times through this temperature regime.  

The product of the first three panels, $n_{\OVI}$, is plotted in the
bottom right panel of Fig. \ref{fig:rvir_OVI}.  The highest $\OVI$
densities are associated with $L^*$ galaxies, with $n_{\OVI}$ peaking
just inside $R_{200}$.  The $\OVI$ mass as a function of radius peaks
outside $R_{200}$, since although the density is lower, the volume
increases as $4 \pi r^2$.  

\subsection{Nature and origin of circumgalactic oxygen} \label{sec:OVInature}

\begin{figure*}
\includegraphics[width=0.49\textwidth]{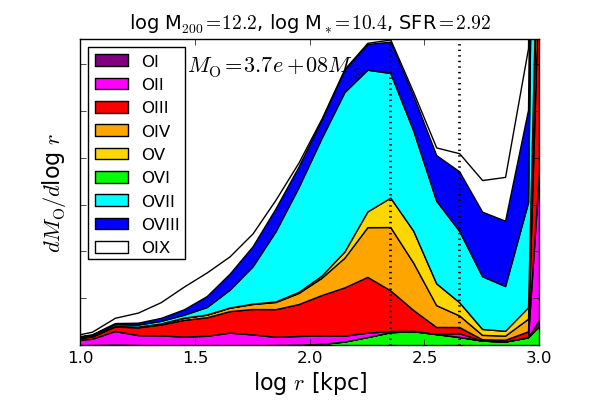}
\includegraphics[width=0.49\textwidth]{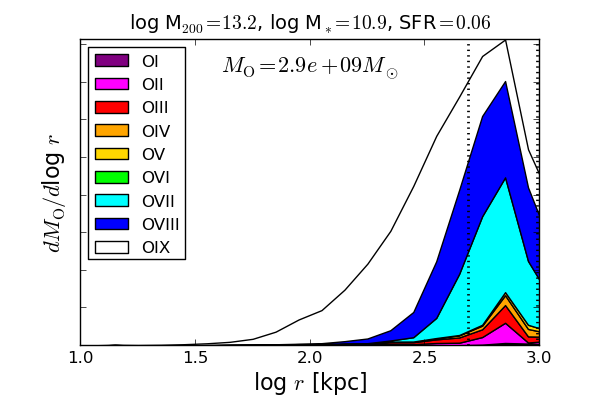}
\includegraphics[width=0.49\textwidth]{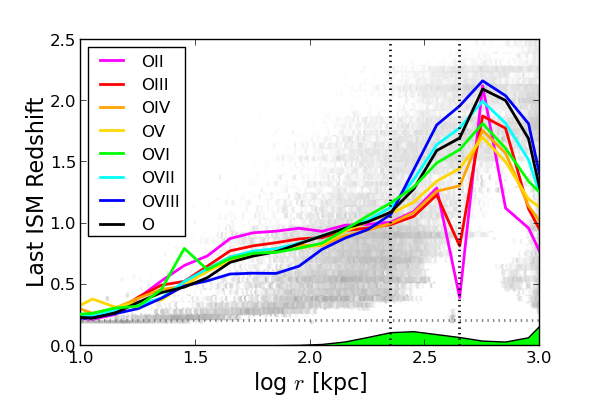}
\includegraphics[width=0.49\textwidth]{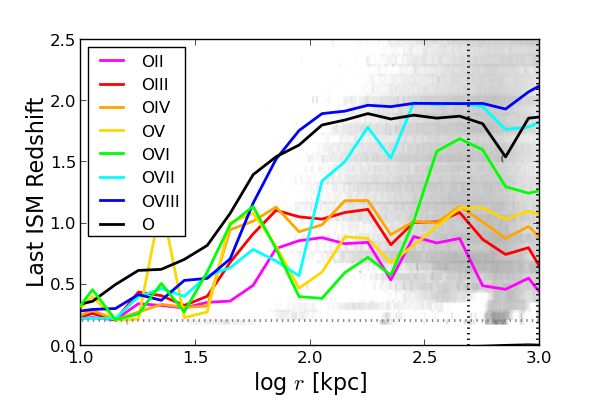}
\caption[]{CGM oxygen as a function of radius is plotted in the top
  panels for the Gal005 $L^*$ halo (left, $M_{200}=10^{12.2} \msolar$)
  and the Grp008 group-sized halo (right,
  $M_{200}=10^{13.2} \msolar$).  The top solid line indicates the
  total oxygen content as a function of radius on a linear scale
  (y-axis values not specified).  The total CGM oxygen content in each
  panel between $10^{1.05}-10^{2.95}$ kpc is $3.7\times10^8$ and
  $2.9\times10^9 \msolar$ for the $L^*$ and group haloes respectively
  (listed on the top of each panel).  The dotted vertical lines
  indicate $R_{200}$ and $2R_{200}$. The bottom panels indicate the
  age distribution of oxygen in gray shading, where we define age as
  the last redshift an SPH particle was in the ISM.  The mean ages of
  oxygen species (coloured lines) and total oxygen (black line) are
  also shown.  The gray horizontal dotted line is the redshift of the
  snapshot.  We also repeat the $\OVI$ shading from the top panels in
  the age panels.  $\OVI$ in $L^*$ haloes primarily resides outside
  $R_{200}$ and is greater than 5 Gyr old ($z>1$).  Group-sized haloes
  enrich a more extended region at earlier times ($z\sim 2$).}
\label{fig:mass_age}
\end{figure*}

Thus far we have focused on the trends of oxygen and $\OVI$ with halo
mass, and we now shift our focus to the distribution and origin of
oxygen-enriched gas in the circumgalactic medium.  We plot the oxygen
content as a function of radius for our standard $L^*$ haloes in the
top left panel of Figure \ref{fig:mass_age}.  Green shading on the
bottom shows the $\OVI$ fraction of the oxygen budget peaking between
one and two $R_{200}$, which are represented by vertical dotted lines.
It is notable that there are five oxygen ions with greater
contributions than $\OVI$, and that $\OVI$ is one of the most
spatially extended ions.  This is the general result for $L^*$ haloes,
$\OVI$ resides primarily beyond $R_{200}$ or at $\sim 200-500$ kpc.
This suggests that a significant fraction of the $\OVI$ detected by
COS-Halos at $b<150$ kpc likely arises at physical radii larger than
$R_{200}$.

In the bottom left panel of Figure \ref{fig:mass_age}, we show the
``age'' of the CGM, by plotting the last time a CGM particle was in
the ISM of a galaxy, which is defined as the last time the gas
particle was star-forming (either in the central galaxy or a
satellite).  The vast majority of metals in the CGM were previously in
the ISM and launched by superwind feedback into the CGM as found in
other SPH simulations \citep{cra13, for14}.  The grey shading shows
the oxygen-weighted age distribution as a function of radii, showing
an age-radius anti-correlation, where larger radii and lower
overdensities are enriched by higher redshift progenitors when halo
potentials are lower and comoving distances are physically smaller
\citep{dav07, wie10, opp12, cra13, for14}.  $\OVI$ has a mean
enrichment redshift of $z\sim 1.2$ at 300 kpc.  Our simulations
suggest that the average age of $\OVI$ is $\sim 6$ Gyr, and it is thus
not associated with the on-going SF traced by observational
indicators, which probe galactic star formation on a much more recent
timescale, i.e.  $\la 100$ Myr.

In the right panels of Fig. \ref{fig:mass_age}, we plot the relative
oxygen content and age distribution around our largest group.  The
$\OVI$ is hardly visible on this linear plot, peaking at nearly $2
R_{200}$.  This group halo is dominated by ions that are only
observable in the X-ray, with $\OIX$ dominating within $R_{200}$.  By
plotting the oxygen distributions on a physical rather than fractional
virial scale, we emphasize the stark contrast between the linear
distributions of oxygen between $L^*$ and group haloes within 1 Mpc.
The progenitors of this central passive galaxy have injected most of
their metals at $z\sim 2$ where they now reside at larger radii
relative to the $L^*$ haloes, both in terms of physical distance and
fractional virial distance.  

\section{Comparison to COS-Halos data} \label{sec:COS-Halos}

We confront the COS-Halos observational survey using our new python
module called Simulation Mocker Of Hubble Absorption-Line
Observational Surveys (SMOHALOS).  We use SMOHALOS to create mock
COS-Halos surveys using observed impact parameters for galaxies with
similar properties as observed by COS-Halos, which in this case are
defined by $M_*$ and sSFR.  This enables us to take the latest dataset
from COS-Halos with updated spectroscopic galaxy observations
\citep[e.g.][]{wer12}, and make a mock survey with matched properties.

We do not constrain our mock galaxies to have the same redshifts as
COS-Halos ($z= 0.15-0.35$).  Instead, we select the {\it M5.3}
simulated galaxy with the most similar $M_*$ and sSFR that are closest
to the observed values from as many as 6 output redshifts ($z=0.250,
0.205, 0.149, 0.099, 0.053, 0.0$) for which we have run our NEQ zooms
for a sufficient time ($t>300$ Myr).  Our galaxies evolve little over
this redshift, except for the cases where there are mergers.  We find
that the $\OVI$ does not evolve significantly over this redshift range
either, which we discuss later.  The main rationale for this choice is
to obtain more passive galaxies that satisfy the isolation criteria,
because neighbours are either beyond 300 kpc at earlier snapshots, or,
more often, merge into a single galaxy to satisfy the isolation
criteria at later snapshots.  Also, we did not output snapshots
between $z=0.25-0.35$.  Figure \ref{fig:abundance_ssfr} shows galaxies
that are isolated in one of three projections as solid circles.  We
have in total 123 $L^*$ and group galaxies across six redshifts and 20
zoom simulations, 87 of which are isolated in all 3 projections, and
108 of which are isolated in at least one projection.  One group halo
is not isolated in any projection (Grp005).

\begin{figure*}
\includegraphics[width=0.49\textwidth]{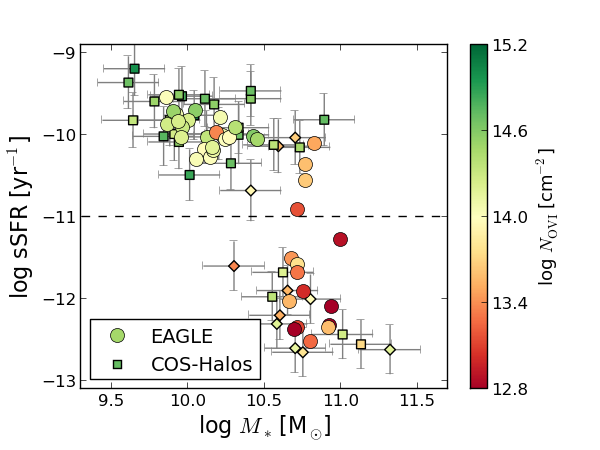}
\includegraphics[width=0.49\textwidth]{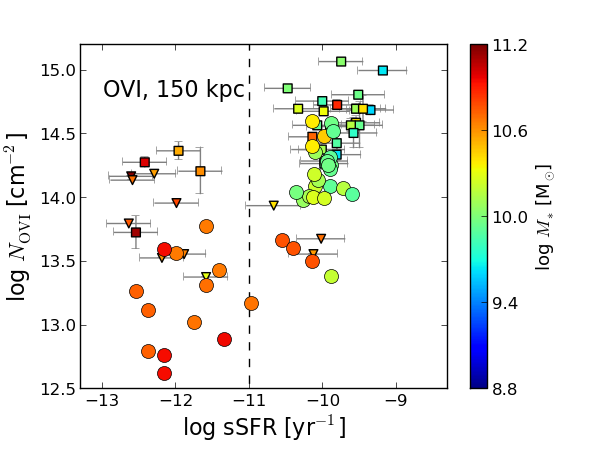}
\includegraphics[width=0.49\textwidth]{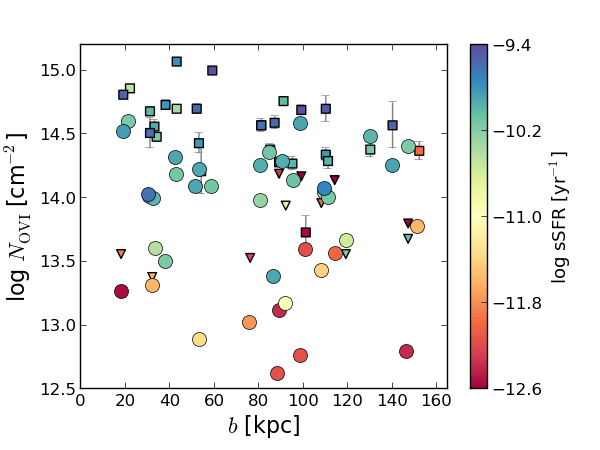}
\includegraphics[width=0.49\textwidth]{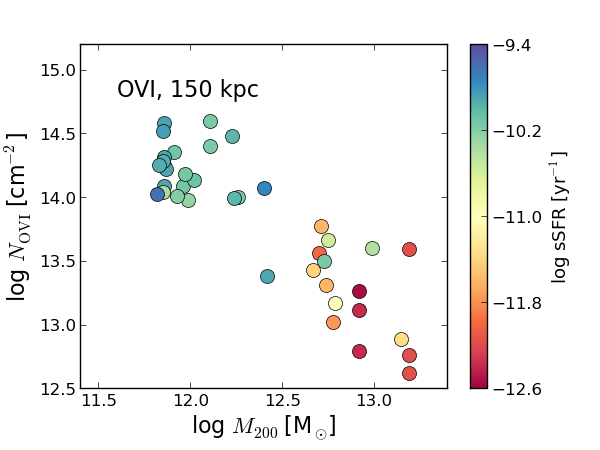}
\caption[]{Comparisons of the COS-Halos dataset (squares are
  detections, triangles pointing down are upper limits) to a mock
  COS-Halos survey (circles).  We choose the simulated galaxy that is
  closest to the given COS-Halos galaxy in terms of $M_*$ and sSFR,
  and choose the same impact parameter, $b$, as the observed galaxy in
  a projection for which the galaxy would be defined as isolated from
  other $L^*$ galaxies.  The top left panel shows real and mock
  observations on the $M_*$-sSFR plane, coloured by $\OVI$ column
  (diamonds here indicate galaxies with $\OVI$ upper limits).  The top
  right panel plots $N_{\OVI}$ as a function of sSFR, coloured by
  $M_*$.  The bottom left panel shows $N_{\OVI}$ as a function of $b$,
  coloured by sSFR.  The bottom right panel shows only the mock
  dataset coloured by sSFR to demonstrate how $\OVI$ columns change as
  a function of halo mass in a mock COS-Halos survey.  Dashed lines in
  the top panels indicate the division between active and passive
  galaxies using a sSFR cutoff of $10^{-11} {\rm yr}^{-1}$. }
\label{fig:COS-Halos_NOVI}
\end{figure*}

We use the T11 list of COS-Halos galaxies, omitting the two least
massive galaxies along each of the J0042-1037, J0820+2334, and
J0943+0531 sight lines ($M_*<10^{9.5}\msolar$), since we do not
explicitly simulate such masses, which leaves 39 galaxies in the
COS-Halos sample.  Stellar masses assume a \citet{cha03} IMF, which
decreases the stellar masses reported by \citet{wer12} by 0.2 dex, who
assumed a \citet{sal55} IMF.

We take one-sigma errors on $M_*$ and sSFR, where we assume
$\delta M_*=0.2$ dex and use the Balmer-derived SFR and error from
\citet{wer12} to calculate the total error on the sSFR, which are
added in quadrature to the error on $M_*$ resulting in
$\delta$sSFR$=0.30-0.37$ dex.  SMOHALOS applies a Gaussian dispersion
seeded by a random number to the values of the simulated galaxies for
the given impact parameter.  We choose the simulated galaxy closest on
the $M_*$-sSFR plane to the observed galaxy given the errors.  We run
100 such realizations of SMOHALOS to construct 100 simulated COS-Halos
galaxy samples using different random number seeds.

Figure \ref{fig:COS-Halos_NOVI}, top left, shows a typical SMOHALOS
mock sample, plotted as circles, on the $M_*$-sSFR plane, with the
COS-Halos galaxies plotted as squares for galaxies with $\OVI$
detections and diamonds for galaxies with $\OVI$ upper limits.  Error
bars indicate uncertainties on observations.  The galaxies are
colour-coded by their $\OVI$ columns in both cases.  It is immediately
apparent that while our galaxies generally reproduce the star-forming
and passive distributions seen by COS-Halos, we do not always obtain
the same range of properties.  For example, we do not create galaxies
with $M_*>10^{11.0} \msolar$, whereas three of these exist in
COS-Halos.

The top right panel shows $\OVI$ column density as a function of sSFR,
coloured by $M_*$, and the bottom left panel shows $\OVI$ columns as a
function of impact parameter, coloured by sSFR (COS-Halos detections
are squares and triangles pointing down are upper limits).  The
COS-Halos trend of $\OVI$ columns increasing with sSFR is reproduced.
However, our mock $\OVI$ columns are too low compared to those
inferred from COS-Halos, both as a function of sSFR and impact
parameter.

We plot the SMOHALOS $\OVI$ columns against halo mass in the bottom
right panel of Fig. \ref{fig:COS-Halos_NOVI} to show that $\OVI$
correlates strongly with $M_{200}$, even more so than with the central
galaxy sSFR (top right panel).  This plot nearly mirrors the trend of
the 150 kpc aperture columns in Fig. \ref{fig:mhalo_NOVI}, in part
because there is not much dependence of $N_{\OVI}$ on impact
parameter.  However, there is greater scatter when choosing pencil
beams corresponding to individual sight lines rather than taking an
averaged aperture column.  The colour-coding of symbols in both plots
indicates that the scatter in group sSFR primarily contributes to the
reduced correlation in the $N_{\OVI}-$sSFR relation relative to that
in the $N_{\OVI}-M_{200}$ relation.  We suggest that the trend of
$N_{\OVI}$ with sSFR found by COS-Halos is driven by halo mass rather
than star formation activity.

\begin{figure}
\includegraphics[width=0.49\textwidth]{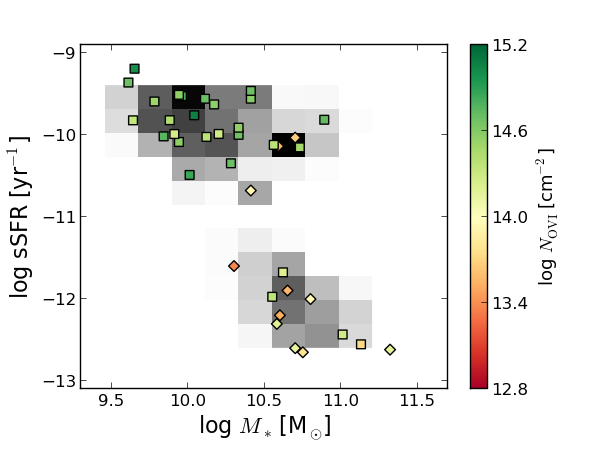}
\includegraphics[width=0.49\textwidth]{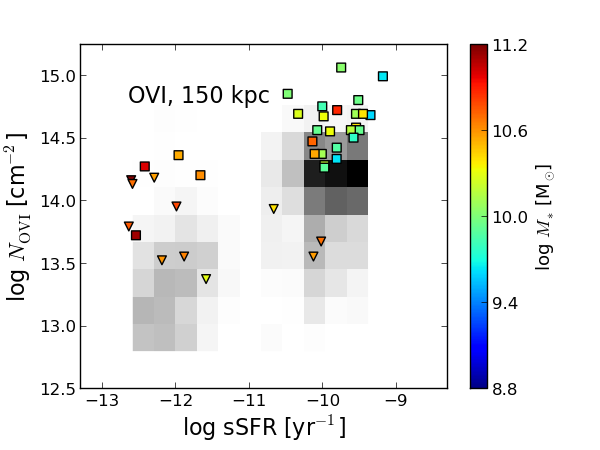}
\includegraphics[width=0.49\textwidth]{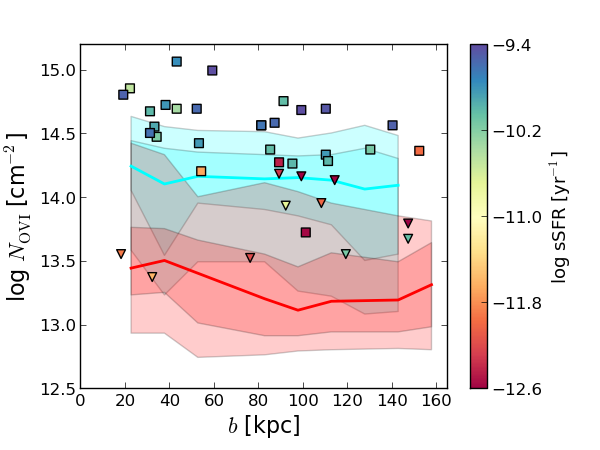}
\caption[]{Probability distributions of mock observations from 100
  SMOHALOS realization displayed in grey shading on a linear scale on
  the $M_*$-sSFR plane (top panel) and the $N_{\OVI}$-sSFR plane
  (centre panel).  COS-Halos data is plotted as in
  Fig. \ref{fig:COS-Halos_NOVI}, but without error bars.  The bottom
  panel shows median $N_{\OIV}$ as a function of impact parameter for
  the star-forming (cyan) and passive (red) SMOHALOS samples,
  including darker and lighter shading for 1 and 2-$\sigma$
  dispersions.  There is hardly any overlap of the 1-$\sigma$
  dispersions between the star-forming and passive samples.}
\label{fig:COS-Halos_prob}
\end{figure}

We plot the probability distributions of the 100 SMOHALOS runs in the
top two panels of Figure \ref{fig:COS-Halos_prob} as grey shading and
superimpose the results of the COS-Halos survey, again as coloured
squares and upper limit triangles.  The top panel shows that we
consistently identify a bimodal set of galaxies in the $M_*$-sSFR
plane.  The middle panel shows that the SMOHALOS mock catalogues
consistently reproduce the $\OVI$-sSFR correlation, although the
column densities are lower than for COS-Halos.  

Sub-dividing galaxies into star-forming and passive samples using an
sSFR cut of $10^{-11}$ yr$^{-1}$, SMOHALOS recovers a median
log[$N_{\OVI}/\cms$]$=14.13$ for the ``blue'' sample (27 galaxies) and
$13.20$ for the passive ``red'' sample (12 galaxies).  The
corresponding COS-Halos numbers are $14.57$ and $<14.14$,
respectively, the latter being an upper limit, owing to most passive
COS-Halos columns being non-detections.  The typical column density of
our blue sample is thus a factor of $2.8\times$ too low, and even if
we take the SMOHALOS sample with the strongest median $\OVI$ column of
the 100 realizations, log[$N_{\OVI}/\cms$]$=14.27$, the column density
is $2.0\times$ too low.

In the bottom panel of Fig. \ref{fig:COS-Halos_prob}, we plot the
average SMOHALOS radial $\OVI$ column density profiles of the
star-forming (cyan) and passive (red) galaxies with shading indicating
1 and 2-$\sigma$ dispersions using the 100 SMOHALOS realizations.  We
rarely reproduce columns in excess of $10^{14.5} \cms$ and never over
$10^{15.0} \cms$ of which there are two detections in COS-Halos.  Also
apparent is the surprisingly flat dependence of $\OVI$ column on
impact parameter for both star-forming and passive samples.  When we
subdivide the blue sample into two impact parameter bins, we find a
larger difference in the interior CGM, such that median
log[$N_{\OVI}/\cms$] for COS-Halos and SMOHALOS are 14.70 and 14.14 at
0-75 kpc, and 14.38 and 14.12 at 75-150 kpc, respectively.  Our median
mass calculation of $\OVI$ within 150 kpc is $1.3\times 10^6 \msolar$,
which is about $2\times$ lower than $2.4\times10^6 \msolar$ calculated
by T11.  We discuss this shortfall of $\OVI$ around star-forming
galaxies in EAGLE in \S\ref{sec:shortfall}.

The interpretation of the comparison to the passive sample of 12
galaxies is more ambiguous owing to the COS-Halos non-detections.  The
median $\OVI$ column density may be consistent between COS-Halos and
the SMOHALOS catalogues; however, we never produce $\OVI$ columns
around passive galaxies in the range of $10^{14.2-14.4} \cms$, as high
as the three strongest COS-Halos passive detections.  We suspect that
our SMOHALOS sample may be cleaner of neighbours than COS-Halos
sample, because observations can miss surrounding galaxies owing to
surface brightness limits.  This was for example the case for the
J2257+1340 sight line where two smaller star-forming galaxies were
subsequently observed closer to a $10^{10.9} \msolar$ passive galaxy
at 114 kpc \citep{tum13}.  Group galaxies with neighbours have higher
aperture column densities, $\langle N_{\OVI} \rangle_{150}=10^{13.76}
\cms$, than for isolated ones, $10^{13.50} \cms$.

\section{The oxygen content of galactic haloes}\label{sec:oxycont}

\begin{figure*}
\includegraphics[width=0.99\textwidth]{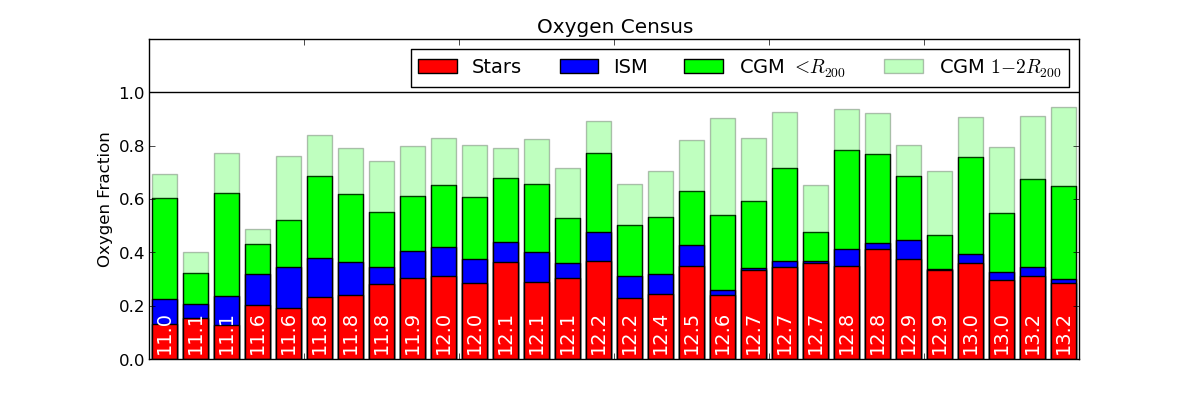}
\caption[]{The oxygen budget for zoom haloes ordered by halo mass
  (log[$M_{200}/\msolar$] indicated by rotated white numbers), where
  the amount of oxygen produced and released by all galaxies within
  $R_{200}$ is normalized to one.  The relative oxygen content in
  stars (red), the ISM (blue), and the CGM (within $R_{200}$, bright
  green; at 1-2$R_{200}$, faded green) is then summed relative to the
  total oxygen production, as in P14.  For $L^*$ and group haloes,
  there exists more oxygen in the CGM within $2 R_{200}$ than in the
  galaxy (stars+ISM) itself.  Up to half the oxygen generated by an
  $L^*$ galaxy can be ejected beyond $R_{200}$.  One difference with
  P14 is that we sum the stellar and ISM oxygen content for all
  (central \& satellite) galaxies in the halo, where P14 only include
  the central galaxy for these quantities.  }
\label{fig:Omass_halo}
\end{figure*}

We now consider the total amount of oxygen formed in galaxies, and the
phases this metal occupies at $z=0.2$.  Following \citet[][hereafter
P14]{pee14}, we normalize the total mass of oxygen nucleosynthesized
and ejected by stars to unity for each of our galaxies, and plot the
fraction of oxygen found in stars (red), the ISM (blue), and the CGM
(green) in Figure \ref{fig:Omass_halo}.  Where P14 plotted a
continuous distribution as a function of stellar mass based on
observed metal abundances scaled to expected nucleosynthetic yields,
we plot bars for our individual galactic haloes and order them by
increasing $M_{200}$, listed in white at the bottom of each bar.  Our
mass distribution is not continuous, and the group-sized haloes are
selected to have a large scatter in the baryon fractions within
$R_{200}$.

We measure that EAGLE oxygen yields are on average 2.7\% of $M_*$ at
$z=0.2$.  Even though EAGLE employs metallicity-dependent yields, this
makes negligible difference for the summed oxygen yield.  This value
is nearly the same as found by P14 who adopted an oxygen yield of
1.5\% for the zero-age main sequence and calculated an average
multiple of 1.8 to account for the fact that $\approx 55\%$ of the
stellar mass survives to $z=0$, based on \citet{lei12} star-formation
histories (i.e. $1.5\%\times1.8=2.7\%$).  It is not surprising that
these two calculations agree, because we are using similar yields
tabulated by \citet{wie09b} as P14, and because we are
self-consistently following the star-formation histories accounting
for stellar evolution\footnote{Oxygen that was first released and then
  trapped in the remnants of another generation is included.}

One difference with respect to P14 is that we plot the total oxygen
generated by all galaxies residing in the galactic halo, instead of
just the central galaxy.  This makes a marginal difference for $L^*$
and sub-$L^*$ galaxies, but the difference is quite noticeable for
group-sized haloes since the stellar mass becomes less dominated by
the group central; less than half the stellar mass resides in the
central for haloes of $M_{200}\ga 10^{13.0} \msolar$.  The ISM oxygen
content in the 4 largest haloes is contributed almost entirely by
satellites.

The stellar oxygen content (red bars) follows the P14 trend with $L^*$
and group haloes having similar oxygen fractions in stars (their
Figure 12), but our sub-$L^*$ galaxies ($M_*\sim 10^9 \msolar$) have
marginally higher stellar oxygen content, which is consistent with the
EAGLE-Recal simulation (S15) exhibiting slightly higher stellar
metallicities content than inferred observationally by \citet{gal05}.
The ISM oxygen content increases at lower mass, owing to higher gas
fractions; the stellar mass-ISM metallicity relationship (i.e. the
MZR) in the EAGLE-Recal prescription reproduces the observed decline
with mass \citep{tre04, zah14}.  Thus far, the agreement is good
between our zooms and the observations, because the EAGLE-Recal
prescription broadly reproduces the observed metal content of
galaxies.

We now report the oxygen contents of the CGM within $R_{200}$ and
additionally between 1 and $2R_{200}$, which are not nearly as
well-constrained by observations (P14).  Our main results are
two-fold: 1) there exists more oxygen in the CGM within 2$R_{200}$
than in the stellar and ISM content of the galaxy, and 2) a
significant fraction, sometimes more than half of the oxygen, is
ejected beyond $R_{200}$ for $L^*$ galaxies.

\begin{figure*}
\includegraphics[width=0.99\textwidth]{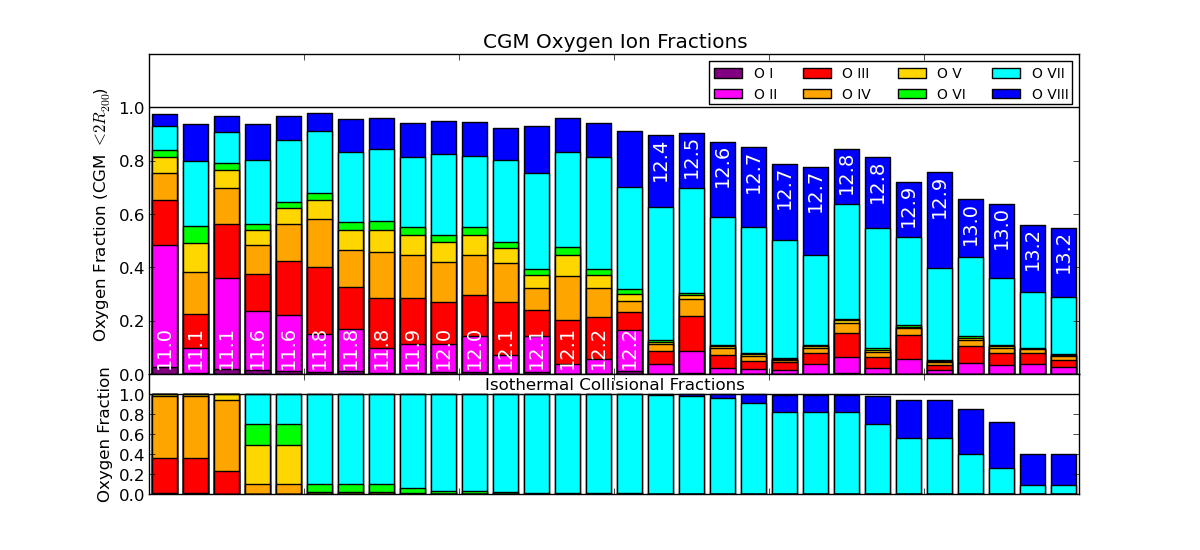}
\caption[]{{\it Top Panel:} Bars indicate the ion fractions of the CGM
  oxygen within $2 R_{200}$ in {\it M5.3} haloes ordered by $M_{200}$.
  This plot normalizes the amount of CGM oxygen represented by the sum
  of the bright and faded green bars in Fig. \ref{fig:Omass_halo} to
  one, and subdivides it into oxygen species.  Eight species
  ($\OI$-$\OVIII$) are coloured, and $\OIX$ is the difference with
  between the end of the blue bar and the horizontal line at a value
  of one.  {\it Bottom Panel:} Bars indicate the ion fractions if one
  assumes an isothermal halo with $T=T_{\rm vir}$ and collisional
  ionization equilibrium.  These assumptions do not hold for simulated
  haloes, but the simulated $\OVI$ fractions transition at a similar
  halo mass, $10^{12.3} \msolar$, as the isothermal fractions
  predict.}
\label{fig:Ofrac_CGM}
\end{figure*}

The first result, that more oxygen exists outside galaxies was also
inferred by P14.  Here, we are omniscient about the location of our
simulated metals, and find that there is almost always more
circumgalactic oxygen within $2 R_{200}$ than in the galaxy
(stars+ISM).  We plot this CGM oxygen within $2 R_{200}$ in Figure
\ref{fig:Ofrac_CGM} categorised by ion.  The eight ions plotted ($\OI$
through $\OVIII$) plus $\OIX$ (not plotted due to its unobservable
nature) sum to unity.  Ions lower than $\OVI$ dominate the CGM
associated with sub-$L^*$ and many $L^*$ galaxies, while ions above
$\OVI$ always dominate the CGM of group haloes, often at the 80-95\%
level indicating the need for a wide-area space-borne telescope
sensitive to soft X-rays to account for the diffuse oxygen around
groups.

Even for $L^*$ haloes, for which $\OVI$ is the strongest, $\OVI$
contributes only a small fraction of the oxygen content of the CGM,
which reflects this ion's maximal $\approx 5\%$ radial ion fraction
(\S\ref{sec:halotrends}).  We plot the CGM within $2 R_{200}$, because
$\OVI$ barely shows up on a plot with the limit of $<R_{200}$.  The
most common ion is $\OVII$, which makes up $23-50\%$ of the
circumgalactic oxygen at $<2 R_{200}$ between
$M_{200}=10^{11.8}-10^{13.0} \msolar$.  It is almost as if nature
conspires to hide diffuse oxygen (and most oxygen generated by stars)
from our telescopes, since $\OVI$ is the most ``easily'' observable
CGM ion.  Nonetheless, $\OVI$ represents a powerful probe that we
argue arises so strongly around $L^*$ galaxies owing to its
temperature tracing the virial temperature of $L^*$ haloes.

In the bottom panel of Fig. \ref{fig:Ofrac_CGM} we plot the oxygen
fractions assuming collisional ionization equilibrium at the virial
temperature $T_{\rm vir}$ according to Equ. \ref{equ:Tvir}.  Our
simulated haloes are of course not isothermal out to $2 R_{200}$, and
show a far greater diversity of oxygen ionization states, which in
part owes to the neglect of photo-ionization in this simplistic
assumption, but most importantly owes to the complex temperature and
density structure of the CGM.  Nonetheless, our simulated $\OVI$
fractions decline at a similar halo mass, $\sim 10^{12.3} \msolar$, as
this simple model predicts, which clearly supports our conclusion that
$\OVI$ traces the virial temperatures of $L^*$ haloes, and not those
of group-sized haloes.  Going to sub-$L^*$ haloes, $\la 10^{11.7}
\msolar$, the simulated haloes do not show a decline in $\OVI$ as
predicted by the isothermal model, because photo-ionization
begins to dominate at these masses.

Hence, our second main result from Figure \ref{fig:Omass_halo} is that
up to half of the oxygen generated by an $L^*$ galaxy resides outside
of $R_{200}$ in the low-$z$ Universe.  The implications for galaxy
formation models are profound.  Not only is the closed-box model of
galaxy growth inaccurate due to continuous accretion onto galaxies,
the idea that galaxies return products of star formation to extended
CGM reservoirs from which they can later re-accrete
\citep[e.g.][]{bow06, opp10, guo11, hen15} may also be incorrect.  The
median oxygen content inside $R_{200}$ for group-sized haloes
($M_{200}=10^{12.7-13.3} \msolar$) is 69\%, which is somewhat higher
than that of $L^*$ haloes ($M_{200}=10^{11.7-12.3} \msolar$) at 62\%.
The increase of metals retained in higher mass haloes should be
expected as larger haloes have deeper potential wells and physically
larger virial radii.  However, the reverse trend may apply because
group-sized haloes have more star formation at early times when the
Universe is physically smaller and halo potentials are shallower,
which allow winds to travel to greater comoving distances.
Additionally, group-sized haloes contain multiple large galaxies with
large satellites more likely to lose their metals from a group halo,
because they currently reside further from the halo centre and
additionally may have fallen in at late times.  Therefore, the
moderate increase in metals retained inside $R_{200}$ from $10^{12}$
to $10^{13} \msolar$ haloes may represent several competing trends.
Finally, the space between the top of the faded green bars and the
black horizontal line in Fig. \ref{fig:Omass_halo} is oxygen ejected
beyond $2 R_{\rm vir}$, which usually resides in the low-density IGM.

\begin{figure*}
\includegraphics[width=0.49\textwidth]{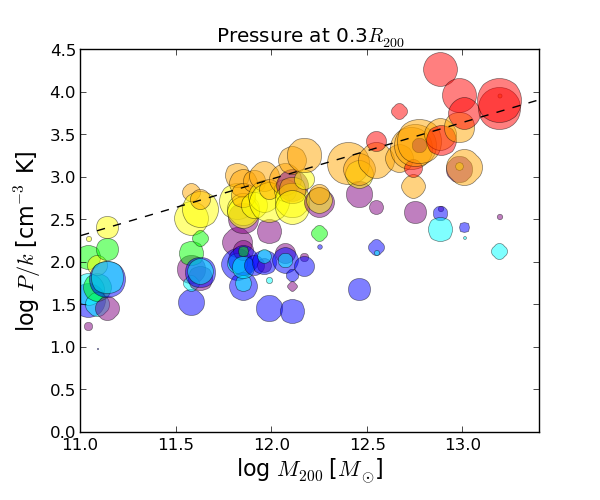}
\includegraphics[width=0.49\textwidth]{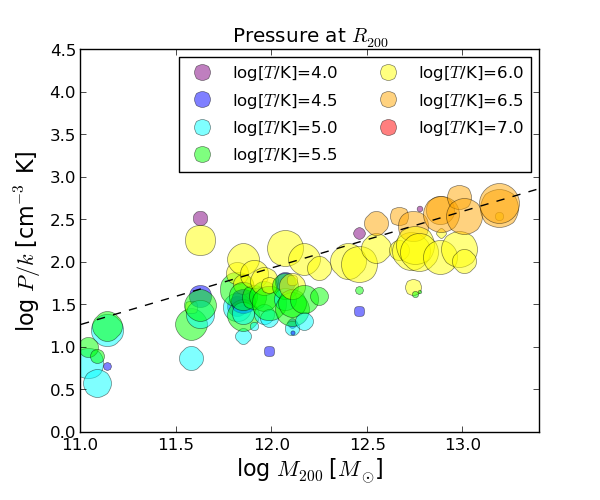}
\caption[]{Pressure as a function of halo mass in shells at
  $0.3\times$ (left panel) and $1.0\times$ $R_{200}$ (right panel).
  Pressures are calculated at a given radius and temperature by
  multiplying the median density within an 0.1 dex spread in $r$ and
  $T$.  The dashed lines indicate the pressure of an isothermal
  density profile at the virial temperature normalized to contain the
  entire baryonic content of a halo within $R_{200}$.  The symbol
  areas scale in proportion to the logarithm of the mass, $m$,
  contained in this phase normalized to halo mass and radius, such
  that area $\propto$ log$[(m/M_{200})(r/R_{200})^{-2}]$.}
\label{fig:Mvir_P}
\end{figure*}

We further investigate the multi-phase structure of the CGM by
considering gas pressures at different temperatures. We plot pressure
as a function of halo mass at 0.3 and 1.0 $\times R_{200}$ in Figure
\ref{fig:Mvir_P}.  To calculate a pressure at a given radius and
temperature, we take the median CGM gas density with a 0.1 dex spread
in radius and temperature, and compute $P/k=\nh T /( X_{\rm H} \mu)$,
where $k$ is the Boltzmann constant, $X_{\rm H}$ is the mass fraction
of hydrogen, and $\mu$ is the mean molecular weight.  For each of our
haloes, we colour the circles by temperature and scale their areas in
proportion to log($m/M_{200}$), where $m$ is the mass within the
temperature/radius bin.

The first point to note is that the CGM is multi-phase, especially
inside $0.3 R_{200}$.  In the $L^*$ range, there are copious amounts
of gas at $T<10^5$ K (purple \& blue circles) and $T\geq 10^6$ K
(yellow, orange, \& red circles).  Very little mass exists at
$10^{5.0}$ and $10^{5.5}$ K (cyan \& green circles), because gas cools
rapidly through this temperature regime.  When the cooling times
become longer at $T\la 10^{4.5}$ K, gas is at a significantly lower
pressure, which indicates this gas is not cooling isobarically.  By
the time the gas reaches $10^4$ K where cooling times are long, it
becomes closer to achieving pressure equilibrium with gas with
100$\times$ more thermal energy.  Thus, the interiors of $L^*$ haloes
inhibit long-lived collisionally ionized, $T\sim10^{5.5}$ K, $\OVI$
owing to rapid cooling through the coronal regime, as well as cool,
$T<10^5$ K, $\OVI$ owing to high pressures forcing cold gas to
densities that are too high to be photo-ionized by the \citet{haa01}
background.  In contrast, at $R_{200}$ where the densities are lower
and the cooling times are longer, the dominant phase is $10^{5.5}$ K
(green circles), where $\OVI$ is a more effective tracer of the gas.

Group haloes have pressures that are far too high to allow long-lived
$\OVI$ at any radii $\la 0.5 R_{200}$, although a multi-phase CGM
exists where a small amount of cold $\sim 10^4$ K gas achieves similar
pressures as much hotter $\sim 10^{6.5}$ K gas in some haloes.  As for
$L^*$ haloes, the gas at $R_{200}$ is less multi-phase, and the
dominant phase is near the $\OVI$ collisional peak temperature.

We overplot the pressures assuming the single-phase isothermal model
profile where $\rho\propto r^{-2}$ and $T=T_{\rm vir}$ from Equ.
\ref{equ:Tvir} with dashed lines in Fig. \ref{fig:Mvir_P}.  The
pressures of the $T>10^{5.5}$ K phase generally follow the approximate
magnitude and the slope ($P\propto M_{200}^{2/3}$) of the model,
although $L^*$ haloes and below have lower pressures interior to $0.3
R_{200}$, which was also shown by \citet{cra10}.  Even though the
single-phase isothermal model is clearly an incomplete description of
the CGM, it provides useful insight into the trends for $\OVI$.  The
high pressures of groups prevent a significant $\OVI$-traced gas
phase.  While $L^*$ haloes have the correct temperature for maximizing
the fraction of collisionally ionized $\OVI$, $\OVI$ still only traces
a significant amount of mass in the exteriors where the coronal
cooling times are long.  Sub-$L^*$ haloes have pressures low enough to
allow photo-ionized $\OVI$.

\section{Discussion}\label{sec:discuss}

Our findings suggest that the $\OVI$-sSFR correlation discovered by
COS-Halos is an indirect correlation, and that the fundamental trend
is $\OVI$ tracing virialized enriched gas at temperatures close to
$10^{5.5}$ K, where the $\OVI$ fraction peaks.  Passive central
galaxies with mass $M_*\sim 10^{10.5} \msolar$ reside in higher mass
haloes than active galaxies with a similar mass.  The higher virial
temperatures of the haloes hosting the passive galaxies lead to
significantly lower $\OVI$ fractions.  While the CGM metal content
increases with halo mass, $\OVI$ is not an effective tracer of haloes
with mass $\gg 10^{12} \msolar$.  The CGM oxygen budget cannot be
directly linked to recent star formation, which we demonstrated by
showing that the typical age of $\OVI$ is greater than 5 Gyr and
represents the cumulative ejection of metals by galactic superwinds
over a Hubble time.

If the $\OVI$ were to trace metals originating from the star formation
determined through observational indicators, then the time to travel
to 150 kpc would imply a very high outflow speed.  A $>$1500 $\kms$
wind is necessary to travel this distance in $<$100 Myr, which is the
timescale for the observed SF indicators to be linked to recent SF.
Such high-velocity winds are rarely if at all present in our simulated
$L^*$ haloes at low-$z$, which is also supported by recent
observations of outflows around star-forming galaxies in the evolved
Universe.  A number of campaigns have observed blue-shifted, low-ion
absorption wings extending to at most a few hundred $\kms$ from the
galaxy, almost always at velocities unable to escape the gravitational
potential of the halo \citep[e.g.][]{bou12, mar12, rub14, chi15,
  nie15}.

We predict the amount of metals residing in low-redshift CGM
reservoirs to be larger than the amount of metals in galaxies
including stars and the ISM.  Despite the high $\OVI$ column densities
around $L^*$ galaxies, the summed $\OVI$ ionization fractions are only
$0.9-1.3\%$ in CGM gas within $R_{200}$ and increase to $2.2-3.4\%$
when including CGM oxygen within $2 R_{200}$.  This is still higher
than the global IGM $\OVI$ correction of $\sim 1\%$ reported by
\citet{rah15b} for the EAGLE 100 Mpc box at low-$z$.  Assuming
$x_{\OVI}=20\%$ thus severely under-estimates the CGM oxygen content.
This is not surprising since collisionally ionized $\OVI$ has a very
small temperature range around $10^{5.5}$ K where it approaches its
maximal $x_{\OVI}$ of $\approx 20\%$.

We emphasize the dramatic decline $\OVI$ takes as one transitions from
$L^*$ to group haloes, and note the strength of the $N_{\OVI}-M_{200}$
correlation. The average $\OVI$ absorption within 150 kpc for {\it
  every} $L^*$ halo in the range of
$10^{11.7}<M_{200}<10^{12.2} \msolar$ is higher than for {\it every}
group halo in the range $10^{12.6}<M_{200}<10^{13.3} \msolar$, where
each subsample contains more than ten haloes.  The $\OVI$ column
densities within 150 kpc decline by a factor of 5 when considering
isolated galaxies, and by a factor of 3.5 when considering all
galaxies, when transitioning from $L^*$ to group haloes.  In contrast,
the oxygen columns within 150 kpc increase by nearly a factor of 3 for
the isolated sample and by more than a factor of 4.5 for all galaxies
moving to higher mass.  Hence, the average $\OVI$ ionization fraction
for the CGM inside $R_{200}$ is almost a factor of 20 lower for
group-sized haloes, such that $x_{\OVI}\sim 0.1\%$.


\subsection{Shortfall of halo $\OVI$ around $L^*$ galaxies}\label{sec:shortfall}

The inability of our simulated ions to reproduce the higher $\OVI$
column densities observed by COS-Halos around blue $L^*$ galaxies is
worth discussing in greater depth.  As noted in \S\ref{sec:COS-Halos},
we find about $2\times$ too little $\OVI$ within 150 kpc using the
{\it M5.3} zooms, with the greatest deviation being within 75 kpc
($\approx 3\times$).  We now discuss several possibilities of why
there may be such a shortfall.

We note that our stellar masses are too low by a factor of $\sim 2$ in
these zooms compared with abundance matching constraints.  In Appendix
\ref{sec:resconv}, we explore the higher resolution {\it M4.4} zooms,
finding that $\OVI$ is slightly stronger ($0.06$ dex) around $L^*$
galaxies, but the stellar masses {\it decline} by 0.18 dex making a
$\ga 3\times$ discrepancy with abundance matching.  Recalibrating the
EAGLE model is necessary at higher resolution, and it may be that
through the generation of more stars, more oxygen is produced and
ejected into the CGM, which creates even higher $\OVI$ column
densities than in the {\it M4.4} zooms.  Indeed, \citet{rah15b} found
that across different EAGLE models, the integrated metal ion column
density distributions scaled nearly linearly with the integrated
stellar mass.  It may also be possible that there exists enough
freedom in superwind feedback implementations to change $\OVI$ columns
by a factor of two, while still successfully reproducing the large
array of galaxy observables at many redshifts as the EAGLE
prescription does.

Another possibility is that the low stellar masses in our {\it M5.3}
zooms could lead to a mismatched selection of over-massive haloes,
where we have shown $\OVI$ to be weaker.  To test this hypothesis, we
run SMOHALOS where we artificially increase the {\it M5.3} stellar
masses by 0.3 dex in accordance with abundance matching constraints,
and find essentially no difference in the median $\OVI$ column
densities.  Hence, this cannot explain the $\OVI$ shortfall.

A third possibility is that the original COS-Halos galaxy-selected
sample exhibits Malmquist bias, which we cannot replicate using
SMOHALOS applied to a limited set of zoom haloes.  The COS-Halos
galaxies were selected photometrically, which can lead to a biased
sample of higher star-forming and/or more massive galaxies from a
Schechter function.  While SMOHALOS attempts to match galaxies based
on $M_*$ and sSFR, we do not have a large enough sample to reproduce
the high SFRs observed in COS-Halos \citep{wer12}.  For example, 3\%
of SMOHALOS galaxies have SFR$>3.0 \msolaryr$ compared to 18\% of
COS-Halos, and we have no galaxies with SFR$>5.0$ $\msolaryr$ compared
to 8\% of the COS-Halos.  Whereas COS-Halos selects star-forming
galaxies from a large cosmic volume, our $L^*$ zooms are selected from
a 25 Mpc box.  Hence, we may not produce the $\OVI$ absorption columns
observed in COS-Halos, because we are not simulating the same
galaxies.

Finally, we assume a slowly evolving extra-galactic background
\citep{haa01} without any local sources of ionizing radiation.
Ionizing radiation from nearby star formation can increase $\OVI$
columns as demonstrated by \citet{vas15} who found that $x_{\OVI}$ can
reach $0.4-0.9$ and reproduce COS-Halos $\OVI$ observations, albeit
under specifically chosen conditions conducive to the photo-ionization
of $\OVI$.  This model could invalidate our conclusion that $\OVI$
traces the virial temperature of $L^*$ haloes and not that of group
haloes, because photo-ionization requires $<10^{5}$ K gas, although it
could be argued that both cold gas and ionizing photons from star
formation decline in group haloes.  \citet{sur16} show that $\OVI$ can
be significantly enhanced by local sources of ionizing radiation, but
only within 50 kpc using a 100\% escape fraction for soft X-rays from
a SFR=10$\msolaryr$ galaxy, which are unrealistic assumptions for
COS-Halos.

However, we suggest the possibility that photo-ionization of $\OVI$
could be part of the solution, owing to fluctuating AGN ``flash''
ionizing their CGM.  \citet{opp13b} introduced the concept of AGN
proximity zone fossils, where ``cool'' $10^{4}$ K gas is ionized by a
halo's central AGN that subsequently turns off leaving the surrounding
CGM over-ionized for a time similar to the recombination timescale of
diffuse gas, which can equal or exceed the typical time between AGN
episodes.  They showed that Seyfert-level AGN that are on for 10\% of
the time can enhance $\OVI$ levels by $>1$ dex after the AGN {\it
  turns off}.  The effect may be particularly relevant for cool metals
at $\nh=10^{-3}-10^{-4} \cmc$ where CGM metals exist in our
simulations, and the relevant recombination timescales for $\OVI$
range from $2-20$ Myrs.  Thus a short Seyfert ``burst'' lasting $\sim
0.1-1$ Myr within the past $\sim 10-20$ Myr can enhance $\OVI$ at the
expense of lower oxygen ions, which normally trace this gas in
equilibrium with the \citet{haa01} field.  The COS-Halos galaxies do
not show AGN activity \citep{wer12} and are thus in the off-phase.
The $\OVI$ nearer to the AGN should be more enhanced by the proximity
zone fossil effect, which would agree with COS-Halos data showing
$\OVI$ increasing at small impact parameter more steeply than our
SMOHALOS realizations.  This hypothesis requires a significant
fraction of COS-Halos blue galaxies to be recent AGN.

\subsection{Effect of AGN Feedback}\label{sec:AGN}

\begin{figure}
\includegraphics[width=0.235\textwidth]{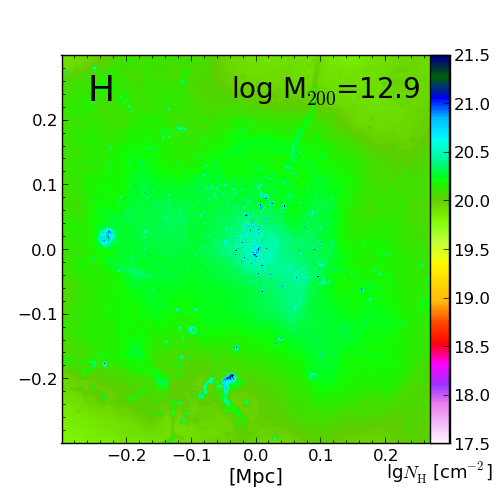}
\includegraphics[width=0.235\textwidth]{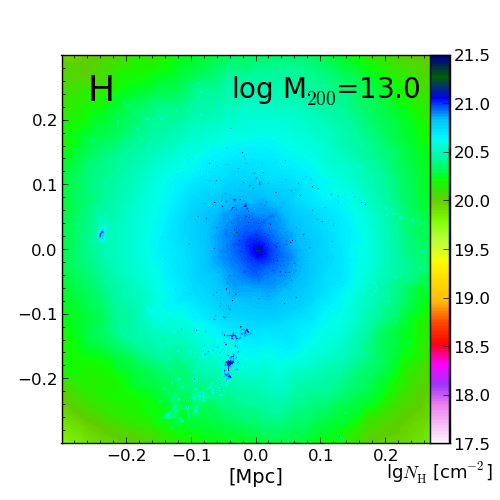}
\includegraphics[width=0.235\textwidth]{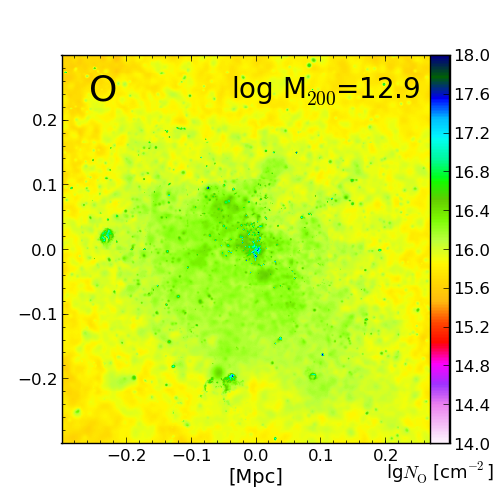}
\includegraphics[width=0.235\textwidth]{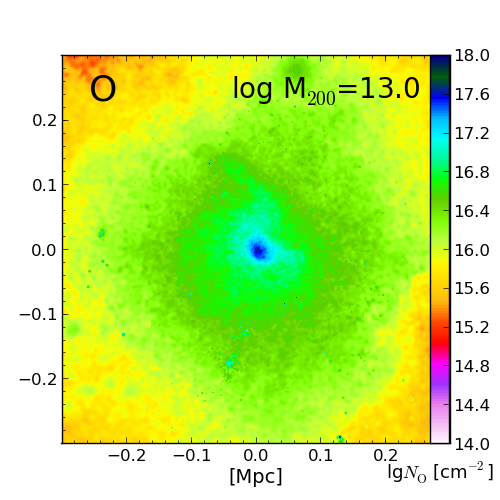}
\includegraphics[width=0.235\textwidth]{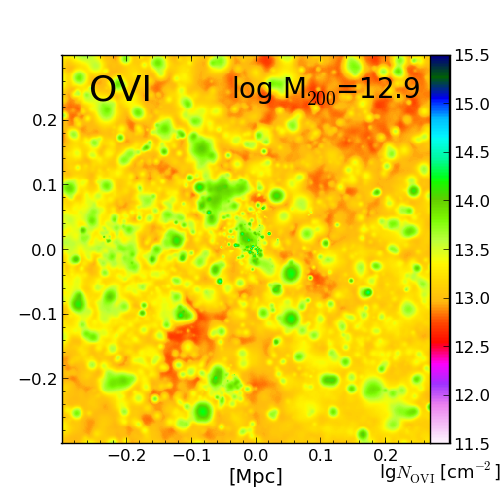}
\includegraphics[width=0.235\textwidth]{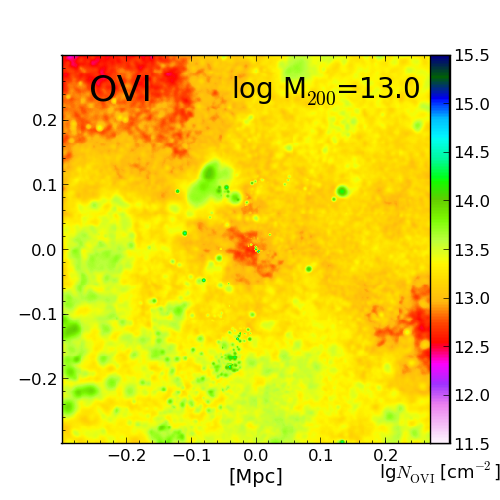}
\includegraphics[width=0.235\textwidth]{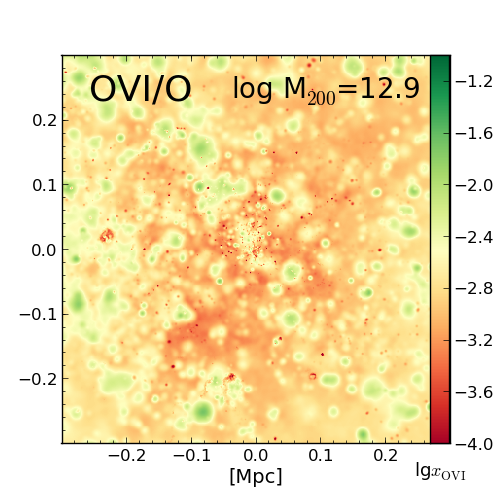}
\includegraphics[width=0.235\textwidth]{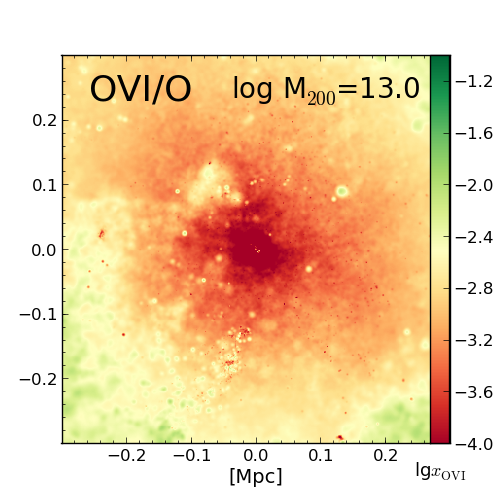}
\caption[]{The same group-sized halo (Grp004,
  $M_{200}=10^{12.89} \msolar$) at $z=0.2$ simulated using the
  standard EAGLE prescription (left) and without AGN feedback model
  (right) in the same format as Fig. \ref{fig:halotrends}.  From
  top-to-bottom, 600 kpc snapshots of hydrogen, oxygen, and $\OVI$
  columns are shown as well as the $\OVI$ ionization fraction.  The
  NoAGN halo has $M_{200}=10^{12.97} \msolar$, compared to
  $10^{12.89} \msolar$ with AGN feedback due to weaker feedback, but
  we use a constant $R_{200}= 387$ kpc when comparing virial
  quantities.}
\label{fig:AGN_noAGN}
\end{figure}

To test the impact of AGN thermal feedback (i.e. not the proximity
fossil effect discussed above) on CGM $\OVI$, we run a single group
halo (Grp004), which has $M_{200}=10^{12.89} \msolar$ at $z=0.205$,
turning off SMBH growth and AGN feedback.  This halo is the most
frequently selected SMOHALOS passive galaxy.  Figure
\ref{fig:AGN_noAGN} compares 600 kpc snapshots of the standard run on
the left versus the ``NoAGN'' model on the right in the same sequence
as Fig. \ref{fig:halotrends}.  While stellar feedback still operates
in the NoAGN run, the galaxy is $2.8\times$ more massive.  There
exists $2.8\times$ more hydrogen and $2.7\times$ more oxygen in the
CGM inside $R_{200}$ in the NoAGN model (cf. top two sets of panels).
The values of $\langle N_{\Oxy} \rangle_{150}$ are $10^{16.19}$ and
$10^{16.65} \cms$ with and without AGN respectively, reflecting the
nearly factor of 3 increase in CGM oxygen in NoAGN group haloes.

Considering the $\OVI$ column densities in the third set of panels,
there exists $1.9\times$ more $\OVI$ within $R_{200}$, but there is
substantially less ($8\times$) $\OVI$ within a radial distance of 100
kpc in the NoAGN model.  The values of $\langle N_{\OVI}
\rangle_{150}$ happen to be very similar, $10^{13.26}$ and $10^{13.35}
\cms$ with and without AGN, respectively, which are both typical
$\OVI$ columns for our simulated groups.  Combined with the oxygen
aperture columns, we show $x_{\OVI}$ is much lower in the NoAGN model
(bottom panels), especially at small impact parameters.  Our findings
for this halo agree with the global IGM statistics of \citet{rah15b}
showing only small differences between the Ref and NoAGN models, which
we further discuss in Appendix \ref{sec:resconv}.

In contrast to \citet{sur16}, who argue that the decline of $\OVI$
around passive versus star-forming galaxies of the same stellar mass
owes primarily to AGN feedback removing mass from the haloes of passive
galaxies, our exploration indicates that it is primarily an ionization
effect that reduces $\OVI$ columns.  Passive COS-Halos galaxies reside
in higher mass haloes than their active counterparts, which have
virial temperatures that are too high for $\OVI$.  Our NoAGN halo has
almost $3\times$ as many stars in the central galaxy and CGM metals
within $R_{200}$ as the standard run, but, almost coincidentally, a
similar amount of $\OVI$ owing to the lower $x_{\OVI}$ as a result of
having less $10^{5-6}$ K gas in the inner halo.  Another big
difference is that the NoAGN run has a SFR$=6.8 \msolaryr$, which
creates a type of isolated galaxy rarely observed in the $z\la 0.2$
Universe.

\section{Summary} \label{sec:summary}

We have run a set of EAGLE zoom simulations to confront the $\OVI$
observations within 150 kpc of $z=0.15-0.35$ star-forming and passive
galaxies from the \citet{tum11} COS-Halos observational campaign.  Our
main simulations use the same model and resolution as the EAGLE
Recal-L025N0752 run, but include the non-equilibrium ionization and
cooling module introduced by \citet{opp13a} and \citet{ric14a}, which
we activate at low redshifts.  Twenty zoom simulations, focusing on
$L^*$ haloes ($M_{200}=10^{11.7}-10^{12.3} \msolar$) hosting
star-forming galaxies, and group-sized haloes
($M_{200}=10^{12.7}-10^{13.3} \msolar$) mostly hosting passive
galaxies, are simulated to $z=0$ with the NEQ module activated at
$z=0.50$ and $0.28$ for the $L^*$ and group haloes, respectively.  Our
main results are as follows:

\begin{itemize} 

\item{Our simulations reproduce the observed correlation of
  circumgalactic $\OVI$ column density and
  sSFR (Figs. \ref{fig:COS-Halos_NOVI}, \ref{fig:COS-Halos_prob})}. 

\item{The bimodality of the $\OVI$ column density, with higher column
  densities around star-forming galaxies than around passive galaxies,
  owes to the temperature of the virialized halo gas.  The virial
  temperature of $L^*$ haloes overlaps with the temperature at which
  collisionally ionized $\OVI$ peaks ($T\sim 10^{5.5}$ K), while
  virial temperatures of group-sized haloes are too high to be traced
  by $\OVI$ (e.g. Figs. \ref{fig:mhalo_NOVI}, \ref{fig:mass_age})}.

\item{There is no direct link between star formation in the central
  galaxy and the $\OVI$ out to 150 kpc.  The median age of the oxygen
  traced by $\OVI$ in the CGM is $>5$ Gyrs, and represents the
  accumulation of circumgalactic metals over the entire star formation
  history of the galaxy.}

\item{The global $\OVI$ fractions of our $L^*$ haloes range between
  $x_{\OVI}=0.9-1.3\%$, which means that far more oxygen exists in
  other ionization states, both lower and higher than $\OVI$.  Most
  $\OVI$ observed around $L^*$ galaxies observed at impact parameters
  $b<150$ kpc is at physical separations of $r=200-500$ kpc ($\approx
  1-2 R_{200}$) from the galaxy.  The global $L^*$ $x_{\OVI}$ rises to
  $2.2-3.4\%$ when considering oxygen out to $2\times R_{200}$, owing
  to optimal ionization $\OVI$ conditions at $\approx 1-2
  R_{200}$ (Figs. \ref{fig:mass_age}, \ref{fig:Ofrac_CGM}).}

\item{Group haloes hosting passive galaxies often have global
  $x_{\OVI}\sim 0.01\%$ inside $R_{200}$, and the majority of
  circumgalactic oxygen is completely ionized ($\OIX$).  Group haloes
  have average column densities of $N_{\OVI} = 10^{13.7} \cms$ for
  $b<150$ kpc, which mostly arises from gas beyond $R_{200}$.  This
  column density is $3.5\times$ lower than for $L^*$ haloes
  ($10^{14.2} \cms$).  Applying isolation criteria similar to
  COS-Halos galaxies, the mean $\OVI$ column density around group
  centrals falls to $N_{\OVI}=10^{13.5} \cms$, which is $5\times$
  lower than for isolated $L^*$ haloes.  Despite the low $\OVI$ column
  densities in groups, there exists $3\times$ more oxygen within a 150
  kpc aperture than for $L^*$ haloes. }

\item{Our simulated $L^*$ haloes have $\OVI$ column densities that are
  $\sim 2\times$ too low compared to COS-Halos observations at $75 < b
  < 150$ kpc and $\sim 3.5 \times$ at $b<75$ kpc.  This could be
    related to an underproduction of total oxygen as our $L^*$ stellar
    masses are $2\times$ too low relative to abundance matching
    constraints.  Increased photo-ionization, possibly from proximity
    zone fossils due to fluctuating AGN \citep{opp13b}, are an
    unconsidered source that could increase the $\OVI$ ion fraction.}

\item{The majority of the oxygen produced and released by stars is
  ejected by superwind feedback beyond the optical extent of the
  galaxy, such that more oxygen exists in the CGM within $2 R_{200}$
  than in the galactic component (stars+ISM).  The progenitors of
  $L^*$ galaxies can eject more than half of their oxygen beyond
  $R_{200}$ by $z=0.2$, although the fraction of oxygen remaining
  within $R_{200}$ (stars+ISM+CGM) averages 62\% and 69\% for $L^*$
  and group haloes respectively.  Enrichment at large radii primarily
  occurs at high $z$, especially for group haloes, when gravitational
  potentials were shallower and fixed comoving distances were
  physically smaller (Figs. \ref{fig:mass_age}, \ref{fig:Omass_halo}).}

\item{$\OVII$ is the most abundant oxygen ion in the CGM for haloes
  between $M_{200}=10^{11.8}-10^{13.0} \msolar$, with between 23 and
  50\% of the circumgalactic oxygen inside $2\times R_{200}$ ionized
  to $\OVII$.  This is a factor of $8-20$ ($60-190$) higher than
  $\OVI$ in $L^*$ (group) haloes (Fig. \ref{fig:Ofrac_CGM}).}

\item{We explore the role of AGN by running a $10^{12.9} \msolar$
  group halo without AGN feedback, finding that it produces similar
  $\OVI$ column densities at $b<150$ kpc ($10^{13.3} \cms$) as our
  standard group haloes with AGN feedback.  Hence, the bimodality of
  circumgalactic $\OVI$ column densities with sSFR does not owe to the
  removal of mass and metals from the haloes of galaxies with more
  active black holes, but is clearly an ionization effect of reduced
  $x_{\OVI}$ in hot haloes (Fig. \ref{fig:AGN_noAGN}).}

\end{itemize}

In our simulations, the higher $\OVI$ column densities around
star-forming galaxies compared with passive galaxies are due to their
virial temperatures being more favorable for a high $\OVI$ fraction
rather than to higher total oxygen column densities.  A key test of
this explanation would be to observe our prediction of the greater
total oxygen content in the CGM of higher mass haloes, where $\OVI$ is
no longer an effective tracer of CGM oxygen.  Observing the soft X-ray
ions $\OVII$ and $\OVIII$ out to the virial radius is beyond the scope
of current facilities, although we will make predictions for upcoming
facilities in future work.  Unfortunately, absorption lines capable of
tracing $>10^6$ K metals are at extreme UV energies, which shifts them
out of the COS wavelength coverage at low redshift.  

Our model contrasts with that of \citet{sur16}, who predict that AGN
blow more baryons out of haloes around passive than star-forming
galaxies, and therefore will not predict more metals at $b \la 150$
kpc around passive galaxies.  Alternatively, photo-ionization by
galactic sources \citep[e.g.][]{opp13b, vas15} instead predicts that
the $\OVI$ bearing material would exist in $\sim 10^4$ K gas around
galaxies.  

Our results indicate that $\OVI$ acts as a sensitive thermometer of
halo gas, such that the $\OVI$ column density around star-forming and
passive galaxies informs more about temperature than about the oxygen
mass.

\section*{acknowledgments}

The authors would like to thank John Stocke, Mike Shull, Molly
Peeples, Romeel Dav\'{e}, Todd Tripp, Kristian Finlator, Brent Groves,
Andreas Pawlik for useful discussions contributing to this manuscript.
We acknowledge Adrian Jenkins for providing the software for the
generation of initial conditions.  We also wish to thank the anonymous
referee for improving the quality of the manuscript.  Support for
Oppenheimer was provided by NASA through grants HST-AR-12841 from the
Space Telescope Science Institute, which is operated by the
Association of Universities for Research in Astronomy, Incorporated,
under NASA contract NAS5-26555, and the Astrophysics Theory Grant,
14-ATP14-0142.  We acknowledge PRACE for awarding us access to
resource Supermuc based in Germany at LRZ Garching (proposal number
2013091919).  This work also utilized the Janus supercomputer, which
is supported by the National Science Foundation (award number
CNS-0821794), the University of Colorado Boulder, the University of
Colorado Denver, and the National Center for Atmospheric Research. The
Janus supercomputer is operated by the University of Colorado Boulder.
This work additionally used the DiRAC Data Centric system at Durham
University, operated by the Institute for Computational Cosmology on
behalf of the STFC DiRAC HPC Facility (www.dirac.ac.uk). This
equipment was funded by BIS National E-infrastructure capital grant
ST/K00042X/1, STFC capital grants ST/H008519/1 and ST/K00087X/1, STFC
DiRAC Operations grant ST/K003267/1 and Durham University. DiRAC is
part of the National E-Infrastructure.  The research was supported in
part by the European Research Council under the European Union's
Seventh Framework Programme (FP7/2007-2013) ERC Grant agreement
278594-GasAroundGalaxies.  Support was also provided by the
Interuniversity Attraction Poles Programme initiated by the Belgian
Science Policy Office ([AP P7/08 CHARM]), as well as the consolidated
grant from the STFC (ST/L00075X/1).  RAC is a Royal Society University
Research Fellow.  AJR is supported by the Lindheimer Fellowship at
Northwestern University.  A database with many of the galaxy
properties in EAGLE is publicly available and described in McAlpine et
al. (2016).

\appendix

\section{Non-equilibrium effects}\label{sec:NEQ}

\begin{figure}
\includegraphics[width=0.49\textwidth]{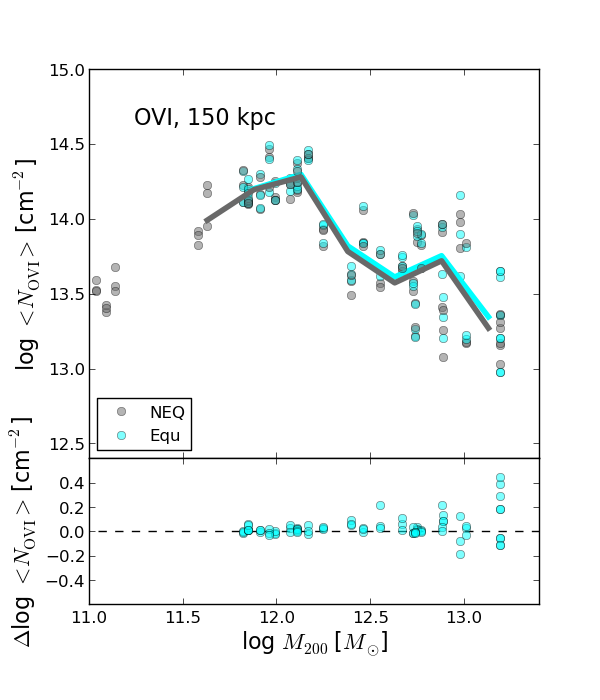}
\caption[]{$\OVI$ aperture column densities within 150 kpc for our
  standard NEQ runs at $z=0.205$ (grey, a repeat of the left panel of
  Fig. \ref{fig:mhalo_NOVI}), and runs with equilibrium ionization and
  cooling (cyan) with the differences relative to the standard NEQ
  runs shown on the bottom panel (for each halo there are 3 circles
  corresponding to 3 projections).  Thick lines of the same colour in
  the top panel are binned averages.  The runs diverge at $z=0.503$
  for $L^*$ haloes and $z=0.282$ for all but one group halo when the
  NEQ module is activated for the NEQ runs. Except for some high-mass
  haloes, the differences are small.}
\label{fig:Equ_OVI_comp}
\end{figure}

Non-equilibrium effects are generally small when considering
circumgalactic $\OVI$ within 150 kpc exposed to a slowly evolving
ionization background such as that specified by \citet{haa01}.  Figure
\ref{fig:Equ_OVI_comp} shows the $\OVI$ column density in the
equilibrium runs (cyan) and the NEQ runs (grey) at $z=0.205$.  The
bottom panel shows that the differences are small-- equilibrium zooms
have $+0.01$ and $+0.02$ dex greater columns than NEQ zooms for $L^*$
and group haloes respectively.  Column densities of high oxygen ions
($\OVI$ and above) typically differ by less than 30\% (0.1 dex) along
individual lines of sight.  The scatter is larger for group haloes
with smaller $\OVI$ columns, where an episodic fluctuation can make a
larger difference.  Galaxy properties ($M^*$ and sSFR) are
indistinguishable between the two sets of simulations.

Our results appear to agree with \citet{cen06b} and \citet{yos06}, who
find that non-equilibrium effects matter for $\OVI$ mainly in the
low-density, diffuse IGM where collisional ionization timescales
become long compared to the timescales on which gas is shocked.  One
caveat is that we may underestimate the NEQ effects, because SPH
smooths out shocks \citep{hut00, cre11}.  Our simulations find that
non-equilibrium ionization and cooling under a slowly evolving
ionization background such as \citet{haa01} do not significantly alter
the $\OVI$ CGM statistics, and will further explore non-equilibrium
effects in such simulations in a future paper.  However, a fluctuating
local source of radiation, as expected for AGN, will introduce
significant NEQ effects \citep{opp13b}.  

\section{Resolution tests and comparisons with large volumes}\label{sec:resconv}

\begin{figure}
\includegraphics[width=0.49\textwidth]{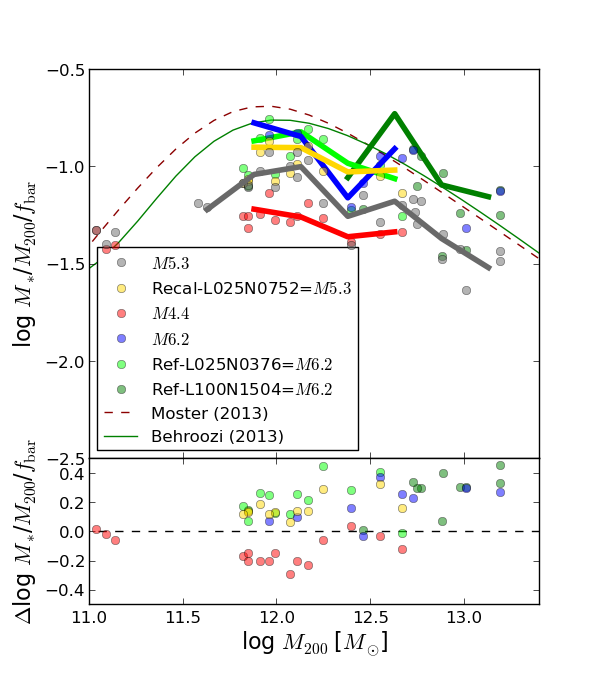}
\caption[]{Galaxy formation efficiency as a function of halo mass for
  our standard NEQ runs at $z=0.205$ (in grey) and for matched haloes
  in Recal-L025N0752 (yellow) at the same resolution ({\it M5.3}), the
  higher resolution {\it M4.4} zooms (red), and the lower resolution
  {\it M6.2} zooms (blue).  We also compare to the matched haloes in
  Ref-L025N0376 (green) and Ref-L100N1504 (dark green).  The outputs
  for EAGLE volumes are at $z=0.187$, and the abundance matching
  constraints from \citet[][solid green]{beh13a} and \citet[][dashed
    red]{mos13} are at $z=0.2$.  The top panel shows the absolute
  values (circles) and binned averages (thick lines), and the bottom
  panel shows differences relative to the {\it M5.3} zooms (grey
  circles in top panel).  }
\label{fig:res_fstar_comp}
\end{figure}

\begin{figure}
\includegraphics[width=0.49\textwidth]{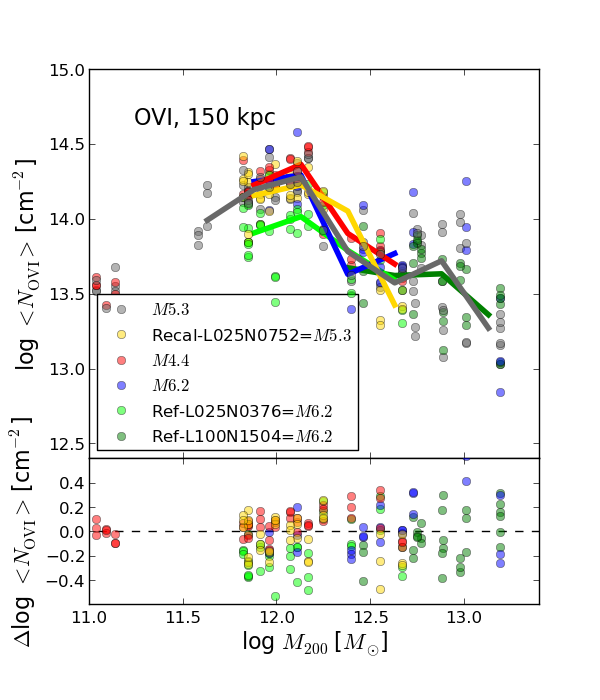}
\caption[]{$\OVI$ aperture column densities within 150 kpc for our
  standard NEQ runs at $z=0.205$ for the same runs as shown in
  Fig. \ref{fig:res_fstar_comp}, showing a range of $64$ ($4$) in mass
  (spatial) resolution.  The aperture is centred on the galaxy
  coordinates given in each respective simulation (zoom and EAGLE
  volume).  The top panel shows the absolute values (circles) and
  binned averages (thick lines), and the bottom panel shows
  differences relative to the {\it M5.3} zooms (grey circles in top
  panel). }
\label{fig:res_OVI_comp}
\end{figure}

We compare our results to higher and lower resolution simulations,
using both our zooms and other periodic EAGLE runs.  We consider
stellar masses relative to abundance matching constraints in Figure
\ref{fig:res_fstar_comp} and 150 kpc aperture $\OVI$ column densities
in Figure \ref{fig:res_OVI_comp}.

\subsection{Metal smoothing at the same resolution}

We first compare matched haloes at the same resolution in the {\it
  M5.3} $L^*$ zooms (grey) and in Recal-L025N0752 (yellow, the
difference with the grey circles indicated in the bottom panel).  This
comparison tests the effect of SPH-smoothing of elemental abundances,
which is done in Recal-L025N0752 but not in the zooms as discussed in
\S\ref{sec:runequil}.  The other differences between these two runs
are NEQ versus equilibrium, which was covered in the previous Appendix
and is insignificant for $M_*$ and $\langle N_{\OVI} \rangle_{150}$,
as well as the zoom method.  We find $0.15$ dex higher stellar masses
for $L^*$ galaxies in Recal-L025N0752, which closely matches the
$0.11$ dex higher stellar masses we find in five {\it M5.3}
smoothed-metallicity zooms.  The $\OVI$ columns are essentially
unchanged, being $0.03$ dex lower in Recal-L025N0752.  Hence, metal
smoothing slightly increases stellar masses but barely alters $\OVI$
CGM columns.  This test also shows that our zooms do not statistically
deviate from their parent simulations, so that our zoom results also
hold for the EAGLE volumes.  Therefore, we can compare our results
directly to \citet{rah15b} for global IGM metal statistics when they
use Recal-L025N0752.

\subsection{Resolution tests}

We have run all {\it M5.3} zooms from the Recal-L025N0752 box at {\it
  M4.4} resolution using the same parameters (Recal prescription and
particle-metallicities; red), which we show in both
Figs. \ref{fig:res_fstar_comp} and \ref{fig:res_OVI_comp}.  These
runs, listed in Table \ref{tab:zooms_app}, constitute a strong
convergence test (S15), and better agreement is potentially achievable
with recalibrated parameters.  Seven out of 10 {\it M4.4} zooms are
run in NEQ from $z=0.503$ to at least $z=0.205$, but the NEQ results
are nearly identical to the equilibrium one.  The large Gal010 zoom
was run in equilibrium only, and provides 3 individual
$M_{200}=10^{12.4}-10^{12.7} \msolar$ $z=0.205$ haloes to test if the
decline in $\OVI$ above $L^*$ persists at this higher resolution.  We
also ran a set of {\it M6.2} zooms using Recal and
particle-metallicities (blue) for nine galaxies in four zooms, also
listed in Table \ref{tab:zooms_app}.

\begin{table*} \label{tab:zooms_app}
\caption{{\it M4.4} and {\it M6.2} zoom simulation runs}
\begin{tabular}{ccrrrrrrrl}
\hline
Name & 
Resolution &
log $M_{200}$ ($\msolar$) &
$m_{\rm SPH}$ ($\msolar$) &  
$m_{\rm DM1}$ ($\msolar$) &
$\epsilon$ (pc) &
log $M_{*}$ ($\msolar$) &
SFR ($\msolaryr$) &
$z_{\rm NEQ}$ & 
$z_{\rm low}$
\\
\hline
\multicolumn {8}{c}{}\\
Gal001  & {\it M4.4} & 12.07 & 2.77e+04 & 1.49e+05 & 175 &  9.98 & 1.166 & 0.503 & 0.0\\
Gal002  & {\it M4.4} & 12.27 & 2.85e+04 & 1.53e+05 & 175 & 10.20 & 1.538 & 0.503 & 0.0\\
Gal003  & {\it M4.4} & 12.11 & 2.80e+04 & 1.50e+05 & 175 & 10.05 & 1.311 & 0.503 & 0.0\\
Gal004 & {\it M4.4} & 12.01 & 2.83e+04 & 1.52e+05 & 175 &  9.93 & 1.077 & 0.503 & 0.0\\
Gal005  & {\it M4.4} & 12.16 & 2.84e+04 & 1.53e+05 & 175 & 10.17 & 1.323 & 0.503 & 0.099\\
Gal006 & {\it M4.4} & 11.93 & 2.77e+04 & 1.49e+05 & 175 &  9.88 & 0.620 & 0.503 & 0.0\\
Gal007 & {\it M4.4} & 11.82 & 2.83e+04 & 1.52e+05 & 175 &  9.76 & 0.953 & 0.503 & 0.0\\
Gal008  & {\it M4.4} & 11.87 & 2.83e+04 & 1.52+05 & 175 &  9.75 & 0.647 & 0.503 & 0.205\\
Gal009  & {\it M4.4} & 11.85 & 2.83e+04 & 1.52e+05 & 175 &  9.79 & 0.772 & 0.503 & 0.205\\
Gal010$^a$  & {\it M4.4} & 12.69 & 2.83e+04 & 1.52e+05 & 175 &  10.55 & 2.019 & -- & --\\
Gal010$^a$  & {\it M6.2} & 12.69 & 1.86e+06 & 9.98e+06 & 700 &  10.93 & 4.677 & -- & --\\
Grp003$^a$  & {\it M6.2} & 12.71 & 1.90e+06 & 1.02e+07 & 700 & 10.99 & 0.708 & -- & --\\
Grp006$^a$  & {\it M6.2} & 12.99 & 1.78e+06 & 9.52e+06 & 700 & 10.87 & 0.000 & -- & --\\
Grp008$^a$  & {\it M6.2} & 13.15 & 1.85e+06 & 9.92+06 & 700 & 11.22 & 0.000 & -- & --\\
\hline
\end{tabular}
\\
\parbox{25cm}{
  $^a$ These haloes were not run in non-equilibrium.\\ 
}
\end{table*}

For $L^*$ haloes, the {\it M4.4} zooms have $0.06$ dex higher $\OVI$
columns despite having $0.18$ dex lower stellar masses as mentioned in
\S\ref{sec:shortfall}.  This means that a greater proportion of the
oxygen formed in higher resolution simulations (since oxygen
production $\propto M_*$) appears as $\OVI$ within 150 kpc of the
galaxy.  The decline of $\OVI$ above $M_{200}>10^{12.3} \msolar$
however is robust with the three Gal010 haloes ($M_{200}=10^{12.42}$,
$10^{12.58}$, and $10^{12.69} \msolar$) showing a considerable decline
in $\OVI$ columns relative to $L^*$ masses.  These three haloes have
$\OVI$ columns averaging $10^{13.79} \cms$, which is $3\times$ lower
than the {\it M4.4} $L^*$ columns-- $10^{14.31} \cms$, and the same
decline as seen in the {\it M5.3} zooms.  Hence, the conclusion that
$\OVI$ traces the virial temperature of $L^*$ haloes and therefore
declines at higher masses is corroborated by these $8\times$ better
mass resolution simulations.  Nevertheless, the {\it M4.4} zooms
underpredict abundance matching constraints by a factor of $\sim 3$,
which implies that the stellar feedback indeed needs to be
recalibrated at {\it M4.4} resolution to inject less energy in order
to create galaxies of the same stellar mass in the $L^*$ mass range.
This is likely a consequence of improved numerical efficiency of our
stochastic thermal feedback implemented at the higher resolution.

The {\it M6.2} zooms are the only galaxies run at this resolution
using the Recal prescription, because L100N1504 uses the Ref parameter
values.  In the $L^*$ range, galaxy stellar masses are on average 0.09
dex higher than {\it M5.3} while the $\OVI$ columns are 0.06 dex
lower.  This continues the trend seen with the {\it M4.4} zooms of a
greater proportion of oxygen produced by stars showing up as $\OVI$
within 150 kpc at higher resolution.  The decline in $\OVI$ above
$M_{200}=10^{12.3} \msolar$ persists at {\it M6.2} resolution.  This
trend is robust at all three resolutions, indicating a decline by a
factor of $\ga 3$ in $\OVI$ columns from $L^*$ to group haloes.  The
resolution convergence is better for $\langle N_{\OVI} \rangle_{150}$
values than for $M_*$ across a factor of $64$ in mass resolution;
however it should be noted that oxygen production is nearly linear
with stellar mass, so that the over-riding trend is a greater fraction
of oxygen produced at higher resolution becomes CGM $\OVI$ within 150
kpc.

\subsection{Comparison to EAGLE volumes}

In the previous section, we compared our results in the ``strong
convergence'' regime, defined by S15 as not changing the subgrid
models across different resolutions, using the Recal prescription.
However, to compare to the main EAGLE simulations that use the ``Ref''
model at {\it M6.2} resolution, we cross-reference the $L^*$
($10^{11.7}-10^{12.3} \msolar$) haloes at {\it M5.3} resolution to
Ref-L025N0376 (green in Figs. \ref{fig:res_fstar_comp} and
\ref{fig:res_OVI_comp}) and the group ($10^{12.7}-10^{13.3} \msolar$)
haloes to Ref-L100N1504 (dark green) to perform a test of ``weak
convergence'' where the subgrid models are changed.  This allows us to
verify our higher resolution zoom results to the much larger sample
from the EAGLE volumes, which are used throughout many EAGLE papers
including \citet{rah15b}.

Relative to our {\it M5.3} zooms, the Ref model has $L^*$ (group)
stellar masses that are 0.21 (0.31) dex higher, while $\OVI$ columns
are 0.24 (0.06) dex lower.  The Ref model better matches the abundance
matching constraints (by design), while our zooms have the additional
difference of using particle rather than smoothed-SPH metallicities on
top of the weak convergence comparison of the standard Ref and Recal
prescriptions.  The decline from $L^*$ to group haloes for
log$[\langle N_{\OVI}\rangle_{150} /\cms]$ is smaller ($13.96$ to
$13.56$) than in the matched {\it M5.3} zooms ($14.20$ to $13.62$),
but still significant.

To generalize these statistics to all haloes in the EAGLE Ref volumes,
we find the average log$[\langle N_{\OVI}\rangle_{150} /\cmc]$ in $L^*$ (group)
haloes using Ref-L025N0376 (Ref-L100N1504) of ($13.83$) $13.61$, where
each sample contains at least 50 haloes.  While the decline is still
apparent, it is smaller at the Ref resolution.  We also used the
intermediate-sized Ref-L050N0752 box to find $13.83$ ($13.66$) for 294
(52) $L^*$ (group) haloes.  However, it must be noted that this sample
includes {\it all} the haloes in 0.6 dex $M_{200}$ bins, which biases
each bin to be dominated by haloes at the lower mass end of the bin
(where there are more haloes) as opposed to the zooms chosen to more
evenly sample the masses within these bins.  Indeed, making the
$M_{200}$ bins smaller (0.3 dex wide) in Ref-L050N0752, leads to a
slightly greater decline from $13.89$ to $13.61$.

The $L^*$ $\OVI$ columns are considerably smaller at Ref resolution,
which is again apparent when we look at all $L^*$ haloes in
Recal-L025N0752 finding $\langle N_{\OVI}\rangle_{150} =10^{14.09} \cms$,
which is 0.26 dex higher than for Ref-L025N0376.  This supports the
findings of \citet{rah15b} where significantly more high $\OVI$ columns
($>10^{14} \cms$) occur in Recal-L025N0752 than in the Ref lower
resolution simulations.  While group halo $\OVI$ columns appear to not
change much in the weak convergence limit, the $\OVI$ columns in $L^*$
haloes are significantly higher in the Recal simulations, which
affects both COS-Halos type observations and global IGM observations
\citep{rah15b}.

\noindent{\bf NoAGN Simulations:} Finally, we consider the
NoAGN-L050N0752 run ({\it M6.2} resolution), and find log[$\langle
  N_{\OVI}\rangle_{150} /\cms$] $=13.96$ ($13.58$) for $L^*$ (group) haloes,
which compares to $13.83$ ($13.66$) in Ref-L050N0752.  Thus, the
conclusion that AGN are not responsible for the reduction of $\OVI$
columns with mass as discussed in \S\ref{sec:AGN}, holds also for this
larger sample of group haloes, and the difference in $\OVI$ column
density is actually {\it stronger} in the absence of AGN.  Globally,
there exists more $\OVI$ in the NoAGN models \citep{rah15b}, which
likely owes to $L^*$ haloes hosting more $\OVI$ while group haloes
harbour less $\OVI$ within 150 kpc despite forming twice as many
stars.

\end{document}